\documentclass[12pt,preprint]{aastex}

\slugcomment{astro-ph/16 Feb. 2004}
\shorttitle{Dust in the Photospheric Environment II. }
\shortauthors{T. Tsuji, T. Nakajima, \& K. Yanagisawa}

\begin{document}

\title{DUST IN THE PHOTOSPHERIC ENVIRONMENT II.
EFFECT ON THE NEAR INFRARED SPECTRA OF L AND T DWARFS
\thanks{Based on the data collected at the Subaru Telescope, which is
operated by the National Astronomical Observatory of Japan}  }

\author{TAKASHI TSUJI}
\affil{Institute of Astronomy, School of Science, The University of Tokyo \\
2-21-1, Osawa, Mitaka, Tokyo, 181-0015, Japan}
\email{ttsuji@ioa.s.u-tokyo.ac.jp}

\author{TADASHI NAKAJIMA}
\affil{National Astronomical Observatory \\
 2-21-1 Osawa, Mitaka, Tokyo, 181-8588, Japan} 
\email{tadashi.nakajima@nao.ac.jp }

\and

\author{KENSHI YANAGISAWA}
\affil{Okayama Astronomical Observatory,
National Astronomical Observatory \\
Kamogata, Okayama, 719-0231 Japan}
\email{yanagi@oao.nao.ac.jp}

\begin{abstract}
We report an attempt to  interpret the spectra of L and T dwarfs
with the use of the Unified Cloudy Model (UCM).
For this purpose, we extend the grid of the UCMs to the cases of
log $g = 4.5$ and 5.5. The dust column density relative to the gas
column density in the observable photosphere is larger at the higher  
gravities, and molecular line intensity is generally smaller at the
higher gravities. The overall spectral energy distributions (SEDs) are 
$ f_{J} < f_{H} < f_{K} $ in middle and late L dwarfs,
$ f_{J} < f_{H} > f_{K} $ in early T dwarfs (L/T transition objects), and
finally $ f_{J} > f_{H} > f_{K} $ in middle and late T dwarfs, where
$ f_{J}, f_{H}$, and $f_{K} $ are the peak fluxes at $J, H,$ and $K$
bands, respectively, in $f_{\nu}$ unit. This
tendency is the opposite to what is expected for the temperature effect, 
but can be accounted for as the effect of thin dust clouds formed deep in
the photosphere together with the effect of the gaseous opacities including
H$_2$ (CIA), H$_2$O, CH$_4$, and K I. Although the UCMs are semi-empirical
models based on a simple assumption that thin dust clouds form in the
region of $ T_{\rm cr} \la T \la T_{\rm cond}$ ($ T_{\rm cr} \approx 1800$\,K
is an only empirical parameter while $ T_{\rm cond} \approx 2000$\,K is
fixed by the thermodynamical data), the major observations including the
overall SEDs as well as the strengths of the major spectral features are
consistently accounted for throughout L and T dwarfs. In view of the 
formidable complexities of the cloud formation, we hope that our UCM can be 
of some use as a guide  for future  modelings of the ultracool dwarfs
as well as for interpretation of observed data of L and T dwarfs. 

\end{abstract}


\section{INTRODUCTION}

So far, few models are available for interpretation and analysis of 
the spectra of L and T dwarfs consistently. Especially, it is well
recognized that dust forms in the photosphere of L dwarfs, but it is
by no means clear how to take the effect of dust into account in the
predictions of the spectra and the spectral energy distributions (SEDs). 
Our initial attempt simply assumed that dust forms everywhere so long
as the thermodynamical condition of condensation is met (Tsuji, Ohnaka, \&
Aoki 1996). Although such models could  explain the spectra of late M dwarfs
and early L dwarfs \citep[e.g.][]{jon97, tsu00, sch01}, at least 
qualitatively, they
failed to explain the spectra of cooler L dwarfs as well as of T dwarfs.
In fact, the photospheres will soon be filled with dust if the simple 
thermochemical equilibrium including condensation is assumed, and
the optical thickness of dust is so large that the predicted spectra
from such a model  will simply be a
blackbody radiation of  $T = T_{\rm eff}$ for $ T_{\rm eff} \la 1500$K or so
(Tsuji 2000, 2001). The fully dusty models by other authors 
\citep[e.g.][]{all01} may have the same difficulty. 
On the other hand, cool T dwarfs, whose prototype is Gl 229B, show no
evidence of dust in their spectra. A naive interpretation was  that the dust 
may have segregated from the gaseous mixture and precipitated below the 
photosphere \citep[][]{tsu96b, mar96, all96, feg96}. However, a question is 
why such  segregation of the dust took place only in cool T dwarfs. 

As a possibility to resolve such difficulties, we proposed a new model
which we referred to as the unified cloudy model, UCM \citep[][]{tsu01} 
and extended it to a grid (log $g$ =5.0 and $ 800 \le T_{\rm eff}
\le 2600$\,K) for applications to L and T dwarfs (Tsuji 2002; hereafter 
referred to as Paper I).  In the UCMs, the segregation of dust from the
gaseous mixture takes place in all the ultracool dwarfs including L and T
dwarfs and at about the same
temperature referred to as the critical temperature $T_{\rm cr}$. Then, 
roughly speaking,  the dust will remain in the observable photosphere 
for the relatively warm dwarfs  with $T_{\rm eff} > T_{\rm cr}$ (note that 
$T \approx T_{\rm eff}$ at $\tau \approx 1$ and hence the region 
of $T \ga T_{\rm cr}$, where dust still survives,  is found in the 
optically thin region), and
hence such warm dwarfs as L dwarfs will appear to be dusty. In the cooler  
dwarfs  with  $T_{\rm eff} < T_{\rm cr}$, on the other hand,  
 the optically thin region (i.e. $\tau < 1$ and hence $ T < T_{\rm eff}$)
will be cooler than  $T_{\rm cr}$ and all the dust grains there
will be segregated and precipitated. For this reason, such cool dwarfs as T 
dwarfs will appear to be dust-free. It is to be noted that this assumption 
behind the UCM  is physically more natural than
 to assume that dust once formed never segregate throughout the photosphere
( namely the fully dusty model of case B) or all the dust grains
segregate as soon as they are formed (i.e.  fully dust-segregated 
model of case C). 

The UCM, however, is by no means a self-consistent theoretical model, but 
rather it is a kind of semi-empirical model at present. It should be 
emphasized, however, that  empirical approach often plays an important 
role in modeling stellar photospheres and atmospheres, even in the more 
simple cases where dust plays no role. For example, empirical models
are still widely used for the solar photosphere, not to speak of
the solar atmosphere (i.e. whole the observable layers including the
photosphere, chromosphere, CO-mosphere, transition layer, corona etc.) for 
which no fully theoretical model may yet exist. 
Once the phase transition occurs in the photosphere, it will introduce
complicated phenomena such as those familiar in the meteorology, 
and it appears to be more difficult to build a fully theoretical model 
from the beginning. Instead, we hope to understand the basic features of the 
dust in L and T dwarfs with the simplest possible semi-empirical model which 
is consistent with the known observations as well as with the basic
physics such as thermodynamics.  We notice that some attempts have been 
made in theoretical modelings of the dust formation in L and T dwarfs 
\citep[e.g.][]{ack01, hel01, mar02, coo03, woi03}, but it is not yet clear 
if they provide consistent interpretation of
the major observations throughout L and T dwarfs.

So far, we have already shown that the UCMs provide reasonable account for 
the L/T transition on the color-magnitude (CM) diagram \citep[][]
{tsu03a} as well as major observations such as infrared colors 
and spectra of ultracool dwarfs throughout L and T dwarfs (Paper I).  
This fact implies that the UCMs  may represent 
the physical structure of L and T dwarfs to some extent.
As a next step in observational tests of the UCMs, we examine if 
the calibrated spectra observed with the Subaru Telescope, as detailed in 
a separate paper (Nakajima, Tsuji, \& Yanagisawa 2004), can be fitted with 
the predicted spectra based on the UCMs.  For this purpose, we first 
discuss some details of the UCMs and extend them to cover the possible 
range of the surface gravities and effective temperatures (Sect.\,2).  
Next, we discuss the dependence of the observable properties on the basic 
stellar parameters (Sect.\,3).  
Then we focus our attention on interpreting the spectral energy
distributions or the spectra of  L, L/T transition objects, and T dwarfs 
based on a single grid of UCMs (Sect.\,4). Although we confirm that the 
observed spectra can reasonably be accounted for by the UCMs, 
many problems remain unsolved before a more detailed  confrontation between 
models and observations can be possible (Sect.\,5).

\section{The Unified Cloudy Models}

\subsection{Dust in the Unified Cloudy Models} 
  
The basic features of the UCMs  are essentially based on a
simple thermodynamical argument.
Namely,  the dust forms near the dust condensation temperature 
$T_{\rm cond}$ as soon as  the thermodynamical condition for condensation 
is met, and dust grows to be as large as the critical radius $r_{\rm cr}$ at 
which the Gibbs free energy of formation attains the maximum. 
Since the Gibbs free energy should decrease in any chemical reaction,
the dust grains smaller than the critical radius $ r_{\rm cr}$
cannot grow larger and are in detailed balance with the ambient gaseous 
molecules by repeating formation and dissociation forever so long as the
thermodynamical condition of condensation is fulfilled. On the other
hand, the grains larger than the critical radius $ r_{\rm cr}$ will grow 
larger and eventually segregate from the gaseous mixture. 
This segregation of dust grains will take place  at a slightly lower 
temperature than the condensation temperature $T_{\rm cond}$ and we 
referred to it as the critical temperature $T_{\rm cr}$. Then,
only small dust grains survive in the photosphere in the temperature 
range of $ T_{\rm cr} \la T \la T_{\rm cond} $ and thus a 
thin  dust cloud is formed. Since  $ T_{\rm cond} \approx 2000$\,K
from the thermochemical data, the dust cloud forms deep in the photosphere 
with $T$ as high as 2000\,K independently of $T_{\rm eff}$. 
As a result, the dust cloud moves from the optically 
thin region in L dwarfs to the deeper optically thick region in T dwarfs. 
This migration of the dust cloud gives a direct effect on the
the CM diagram \citep[][]{tsu03a} and possibly on the observed spectra as well.

The critical radius $ r_{\rm cr}$ is related to the  number of 
monomers $n^{*}$ at which the Gibbs free energy of 
formation $\Delta G(n)$ (eqn.(5) of Paper I) attains the maximum and
$$ n^{*} = \biggl({ {8 \pi a_{0}^2 \sigma} \over { 3kT {\rm In}S } } 
\biggr)^{3}, \eqno(1)  
$$
where $a_{0}$ is the radius of the monomer, $\sigma$ is the surface tension
of the condensed grain, and $S$ is the supersaturation ratio
\footnote{Note that eqn.(1) was given as eqn.(6) in Paper I, but
was misprinted }. Then,
$$  r_{\rm cr} = a_{0}\root 3\of{n^{*}} . \eqno(2) $$
Some physical data on dust grains are summarized in Table 1, and the 
critical radii are estimated on the
assumption of a modest supersaturation ratio of $S = 1.1$. 
It is difficult to know the exact value of $S$,  but this cannot be so large
under the high density of the photosphere of cool dwarfs. The resulting
values of $  r_{\rm cr}$ are 0.01 - 0.02\,$\mu$m and this result is
consistent with the fact that the astronomical grains of about these sizes 
are known. In the following computations, we assume a unique value of the 
grain radius $r = 0.01\,\mu$m, and size distribution is not considered. 
However, so long as the grain sizes are small enough (i.e. $r << \lambda$), 
it is known that\citep[e.g.][]{vdh57}
$$ Q_{\rm abs} \propto r , \eqno(3) $$
and the mass absorption coefficient is  almost 
independent of the grain size (by eqns.(3), (4), (6), (8), and (9)).    

The absorption and scattering cross-sections of a dust grain with radius $r$ 
are:
$$C_{\rm abs}=\pi r^2 Q_{\rm ext} (1-\gamma), \eqno(4) $$
and
$$C_{\rm sca}=\pi r^2 Q_{\rm ext} \gamma, \eqno(5) $$
where $Q_{\rm ext}$ and $\gamma$ are the efficiency factor for extinction
and albedo, reaspectively. These data used in our UCMs are based on the 
optical constants found in the literature referred to in Table\,1 and the 
results are shown in Table\,2. The mass of a dust particle is
$$ w_{\rm dust} = 4 \pi r^3 \rho_{\rm dust} /3, \eqno(6) $$
where $\rho_{\rm dust}$ is the density (specific gravity) of the dust species.
The mass fraction of dust grains in gram of stellar material is
$$ f_{\rm dust} = P_{\rm dust}A_{\rm dust}/q(p_{\rm H}A_{\rm H}
+p_{\rm He}A_{\rm He}+p_{\rm H_2}A_{\rm H_2})  \eqno(7) $$
where  $  P_{\rm dust}$ is the fictitious pressure of the refractory element 
(i.e. Fe, Al, and Si for iron, corundum, and enstatite, respectively) that 
would appear when the dust grains were fully dissolved to the monoatomic gas,
$ A_{\rm dust}$ is the  molecular weight for the chemical 
formula of the dust species, and $q$ is the number of the refractory elements
in the chemical formula (e.g. $q=2$ for Al$_2$O$_3$) ($p's$ and $A's$ are
the partial pressures and molecular weights of the species shown by the
suffix). The number of the dust grains in gram of stellar material is
     $$   n_{\rm dust} = f_{\rm dust}/w_{\rm dust}. \eqno(8)   $$
Then the absorption and scattering coefficients due to the dust species per 
gram of stellar material are:
$$ \kappa_{\rm dust} = C_{\rm abs} n_{\rm dust}, \eqno(9) $$
and 
$$ \sigma_{\rm dust} = C_{\rm sca} n_{\rm dust}, \eqno(10) $$
respectively. In the UCMs, these dust absorption and scattering coefficients 
are added to the continuous absorption and scattering coefficients, 
respectively, but only for the layers with $ T_{\rm cr} \la T \la 
T_{\rm cond} $\footnote{With this simple modification of the extinction
coefficients, any available spectral synthesis code can also be applied to 
the UCMs.}.

\subsection{Revision and Extension of the Unified Cloudy Models}

Within the framework of the classical theory of spectral line formation,
the spectra depend on chemical composition, effective temperature,
surface gravity, and micro-turbulent velocity. As to the chemical composition,
we assume the solar system abundances (Anders \& Grevesse 1989; note
that the iron abundance is based on the meteorite value rather than the
photospheric value), but there was a serious problem in the carbon and
oxygen abundances in our initial version of the UCMs (Paper I).
The situation is much improved  with the latest revisions of the C and O 
abundances (Allene Prieto, Lambert, \& Asplund 2001,2002), and
we have updated our UCMs with the new C \& O abundances (log $A_{\rm C} = 
8.39$ and log $A_{\rm O} = 8.69$ on the scale of log $A_{\rm H} = 12.0$). 

Previously we assumed log $g$ = 5.0 and $v_{\rm micro} = 1$\,km s$^{-1}$ 
throughout, but the effect of the surface gravity should be
examined in the analyses of the observed spectra. 
For this reason, we have extended our grid to include two sequences of UCMs 
with  log\,$g = 4.5$ and 5.5. If the radii of ultracool dwarfs are assumed
to be the Jupiter's radius, the cases of log $g = 4.5, 5.0$, and 5.5
correspond to the masses of 13, 40, and 128\,$M_{\rm Jupiter}$, respectively,
and thus our extended grid may cover the possible range of ultracool
dwarfs. The cases of the lower gravities ( log\,$g$ = 3.0 - 4.0) that may 
cover the contracting phases were discussed before \citep[][]{tsu00}, but 
limited to the extreme cases B and C.

In the UCMs, the critical temperature $ T_{\rm cr}$ is left as a free
parameter, which is to be estimated empirically.
Since $T_{\rm cond} (\approx 2000$\,K)
is fixed by the thermochemical data, the value of $T_{\rm cr}$ is
essentially a measure of the thickness of the dust cloud, which  
should have a direct observable effect. For example, we showed that the red 
limits of the infrared colors are redder for the lower values of 
$ T_{\rm cr}$ (i.e., for the thicker dust cloud).  We analyzed different 
infrared photometric systems such as
the 2MASS\, (Kirkpatrick et al. 1999, 2000), MKO \citep[][]{leg02},
and CIT \citep[][]{dah02} systems (see Tsuji 2001, 2002, and Tsuji \&
Nakajima 2003, respectively), and the results consistently showed
$ T_{\rm cr} \approx 1800$\,K. We confirm that this conclusion
will not be affected by the gravity effect (Sect.\,3.3). 
Although the updated grid of the UCMs with log $g$ = 5.0 are with $ T_{\rm cr} 
= T_{\rm surface}, 1700, 1800, 1850, 1900$\,K, and $ T_{\rm cond}$ 
(i.e., the case of $ T_{\rm cr} = 1850$\,K is added and the 1600\,K case 
of Paper I is removed), we restricted to the case of $ T_{\rm cr} = 1800$\,K 
in the models for log $g$ = 4.5 and 5.5 (and some models of case C for
comparison). The $T_{\rm eff}$ values cover the 
range  between 700 and 2600\,K throughout.

As an example of the effect of the surface gravities on the UCMs,
we compare  the UCMs ($ T_{\rm cr} = 1800$\,K) of $ T_{\rm eff} = 1500$\,K
 for log $g = 4.5, 5.0, $ and 5.5 in Fig.1. 
The gas pressures are higher for the higher gravities as expected, but the 
basic features including the convective structure are essentially the same 
for models with different gravities. The effect of the surface gravities
as well as the effective temperatures on the observable properties are
discussed in Sects.\,3 and 4.

\subsection{Molecular Abundances and Dust Column Densities}

As an example of L dwarfs, the vertical distributions of some molecules
and dust grains are shown for the UCM ($ T_{\rm cr} = 1800$\,K) of 
$ T_{\rm eff} = 1800$\,K and log $g = 5.0$ in Fig.\,2a, in which the abscissa 
shows the logarithms of the partial pressure of molecule and the ordinate 
represents the logarithms of the optical depth defined by the
Rosseland mean opacity, log\,$\tau_{\rm Ross}$ (which is decreasing upward). 
The grain abundance is shown by the fictitious pressure of the refractive
element (e.g. Fe or Al) forming the dust grains referred to as $P_{\rm dust}$ 
in Sect.2.1. In this model, corundum (Al$_2$O$_3$)
condenses at about 1950\,K and iron at about 1850K. But corundum as well as 
iron segregates already at  $ T_{\rm cr} = 1800$\,K in our UCM, and thus the 
iron cloud is quite thin. The geometrical thickness of the
corundum cloud is greater than that of the iron cloud, but its 
effect may be less than the iron cloud because of the lower dust column
density (note that Al is less abundant than Fe by about an order). The 
condensation temperature $T_{\rm cond}$ of enstatite (MgSiO$_3$) is below 
1800\,K and silicate cloud does not appear in this model.
Also, these thin dust clouds can not induce convection such as seen in the 
cooler model to be discussed in the following, and the convection zone 
remains deep in the photosphere below the dust clouds. 
The abundance of the dust grains at the strict thermodynamical equilibrium 
is shown by the dotted line. Presently we are not considering the fate of 
these segregated grains, which, however, result in a drastic decrease of 
FeH, for example. In this model, carbon is still largely in CO although 
an appreciable amount of CH$_4$ is already formed and oxygen is mostly in 
H$_2$O throughout the photosphere. These molecules, both above and below 
the dust clouds, contribute to the spectral line formation (see Fig.\,4), 
since the optical thickness of the clouds is not so large in this model.

As an example of T dwarfs, a similar diagram is shown for the UCM 
($ T_{\rm cr} = 1800$\,K) of $ T_{\rm eff} = 1000$\,K and log $g = 5.0$ in 
Fig.2b. In this cooler model, the gas pressure of the dust forming region is
quite high and the condensation temperatures ($T_{\rm cond}$) of iron and 
corundum are as high as 2200\,K. 
In our simplified assumption of the uniform $ T_{\rm cr} $ value throughout,
the iron and corudum clouds appear to be rather thick. 
Enstatite finally appears but at about 1820\,K and thus silicate cloud
is very thin in this model. All these clouds are already immersed in the 
optically thick region and provide little observable effect. 
One possible effect of these dust clouds, however, is that a new 
convective zone is induced because of the steep temperature gradient due to 
the large dust opacities (Paper I). Without the dust
cloud, the convective zone is situated in the deeper layer, and it should
be emphasized that the convection is induced by the dust clouds and not
the reverse. In this cool and dense model,
carbon is mostly in CH$_4$ rather than in CO. On the other hand, oxygen 
not only remains mostly in H$_2$O but also additional H$_2$O will be
formed by the oxygen released from the dissociation of CO. For this reason,
H$_2$O in Fig.\,2a is less abundant than CO while it is more abundant
than CH$_4$ in Fig.\,2b (Note that the abundance of CO in Fig.\,2a and that
of CH$_4$ in Fig.\,2b  are both equal to the carbon abundance). These 
molecules can now be observed without obscuration by the clouds.  
 
The logarithmic ratio of the mass column density of iron grains against 
that of the total gaseous mixture in the observable photosphere is shown 
in Fig.\,3. For comparison, the logarithm of the mass ratio of Fe to H for 
the composition assumed is  -2.75, and thus maximum of about 10\% 
of Fe in the observable photoshere is in the form of iron grains forming 
the dust cloud in the present UCMs.  
The iron cloud first appears in the models of $ T_{\rm eff} = 2100, 2200,$ 
and 2400\,K for log $g$ = 4.5, 5.0, and 5.5, respectively, and it 
will be immersed below the observable photosphere in the models of
$ T_{\rm eff} \la 1200$\,K for all the log $g$ values. However, this does
not necessarily imply that the effect of dust cloud suddenly disappears at
$ T_{\rm eff} \approx 1200$\,K, because of the large non-greyness of
the opacities. We assumed the limit of the observable photosphere to be at 
$\tau_{\rm Ross} = 3$ where it is still not so opaque in the $J$-band region,
since $\kappa_{\rm Ross}$ is dominated by the H$_2$ collision-induced
absorption (CIA) which is very effective in the $K$-band region. Thus, the 
definition of the dust column density in the observable photosphere cannot 
be very accurate, but Fig.\,3 will give some idea.

\section{PREDICTED PROPERTIES OF THE UNIFIED CLOUDY MODELS}

\subsection{Spectral Line Intensities}
As a guide to interpret the spectra, we evaluate the spectral 
line intensities familiar in the classical stellar spectroscopy 
\citep[e.g.][]{uns55, cay63}. For this purpose, we apply the
method of weighting function and define the spectral line
intensity $\Gamma_{\lambda}$ so that the reduced equivalent width $W/\lambda$
at the weak line limit is given by
$$ {\rm log}(W/\lambda) = {\rm log}gf + {\rm log}\Gamma_{\lambda}(\chi) 
                                               \eqno(11)   $$
with
$$  \Gamma_{\lambda}(\chi) = { {\pi e^2} \over {m c^2} }\lambda
 \int_0^{\infty}P(\tau_{\lambda})G_{\lambda}(\tau_{\lambda})
  { {d\tau_{\lambda}} \over {\kappa_{\lambda}} }  \eqno(12)
$$
where $\kappa_{\lambda}$ includes all the background continuous opacities 
due to ions, atoms, molecules, dust, and quasi-continuous sources (e.g. 
H$_2$ CIA, K I lines), 
$$G_{\lambda}(\tau_{\lambda}) = { {2} \over {F_{\rm cont}(\tau=0)} }
  \int_{\tau_{\lambda}}^{\infty}
  { {dS_{\lambda}(t)} \over {dt} } E_{2}(t)dt  \eqno(13)  $$
is the weighting function ($S_{\lambda}$ is the source function and $E_{2}$
is the integrated exponential function), and
$$ P(\tau_{\lambda}) = { {p_{\rm mol}} \over {P({\rm H})} }
                       { {1} \over {u(T)} } 10^{-\chi \theta}
                       (1 - e^{-hc/\lambda kT}) 
                      { {1} \over {\mu_{\rm H} m_{\rm H}}}   \eqno(14)  
$$
is the number of molecules per gram of stellar material at the fictitious 
lower level with statistical weight unity and with the lower excitation 
potential $\chi$ (in eV). Also, $p_{\rm mol}$ is the partial pressure of
molecule of interest, $P$(H) is the fictitious pressure of the hydrogen 
nuclei, $u(T)$ is the partition function, and $\mu_{\rm H}$ is 
the mean molecular weight with respect to the hydrogen nuclei.

The resulting line intensities for a line with $\chi = 0.0$ eV in the 
H$_2$O 1.4\,$\mu$m bands are shown in Fig.\,4a. If the $gf$-value is known,
the reduced equivalent width at the weak line limit can readily be
given by eqn.(11) with log $\Gamma_{\lambda}(\chi)$ value in Fig.\,4a.
Inspection of Fig.\,4a reveals that the H$_2$O 1.4\,$\mu$m bands are
weaker for the higher gravities at the same $T_{\rm eff}$ and stronger
at lower $T_{\rm eff}$ at the given log $g$ value in general. But the
H$_2$O line intensities show a dip at about $T_{\rm eff} \approx 1600 - 
1700$\,K, and this is due to the effect of the dust extinction which
is the largest at about these $T_{\rm eff}$ values. We also
show the line intensities for a line with $\chi = 0.0$ eV in the 
H$_2$O bands near 3.0\,$\mu$m in Fig.\,4a, in which the dip disappeared.
This is because the dust opacity is no longer so important at 3.0\,$\mu$m
as at 1.4\,$\mu$m.   

The line intensities for a line with $\chi = 0.0$ eV in the CO 2.3\,$\mu$m 
bands strongly depend on the gravities as shown in Fig.4b.
This is due to the effect of H$_2$ CIA which is most effective in the
$K$ band region. Since CIA depends on the square of the density,
CO bands suffer serious weakening at the higher gravities. 
Also, the CO 2.3\,$\mu$m bands show rapid decline at about $T_{\rm eff}
\approx 1500$\,K and this is due to the formation of methane at about this
$T_{\rm eff}$ value. Actually, CO bands may be masked by the stronger
CH$_4$ bands at the cooler $T_{\rm eff}$ values, but such an effect is
not taken into account in the present line intensities, since the
methane bands are not considered as a background opacity in eqn.(12).
In contrast, the line intensities for a line ($\chi = 0.0$ eV) of the CH$_4$ 
bands show rapid increase at about $T_{\rm eff} \approx 1500$\,K.
However, they show only minor dependence on the gravities even though
they should also suffer the effect of H$_2$ CIA. This is because the
CH$_4$ abundance is also highly sensitive to the gravities and
this fact may roughly cancel the effect of the increased CIA at the
higher gravities.  
   
Finally, as an example of refractory molecules, the line intensities for 
a line with $\chi = 0.0$ eV in the FeH 1.1\,$\mu$m bands are shown in
Fig.\,4c. Although FeH almost disappears above the iron cloud, it is quite
abundant to give a large line intensity in L dwarfs because of the presence
of  FeH below the iron cloud (Fig.\,2a). As the iron cloud migrates to the
deeper layer with decreasing $T_{\rm eff}$,  all the iron-bearing molecules
are swept by the iron cloud and little FeH  is left in the observable 
photosphere above the cloud. Nevertheless, the non-zero line intensities 
of FeH is found in the models of $T_{\rm eff} \la 1500$\,K (Fig.\,4c), and 
this  is due to a small amount of FeH above the iron cloud which is in 
equilibrium with the solid iron (Fig.\,2b). This small amount of FeH may 
not be sufficient to explain the FeH bands detected in T dwarfs \citep[][]
{bur02b, nak04}. Moreover, this small amount of FeH was assumed to be
in chemical equilibrium with the iron grains which, however, are assumed to 
be precipitated already below the photosphere in our UCMs.  Thus, this small 
amount of FeH is not in chemical equilibrium with dust and further analysis
should be needed to know the effect of the dust grains assumed to have
precipitated in our UCMs. Anyhow, some other mechanism(s) must be considered 
to explain the FeH bands observed in T dwarfs. One possibility may be a 
convective dredge-up of FeH abundant below the clouds by the second 
convective zone (Fig.\,1). Such a possibility was also considered but 
dismissed by Burgasser et al. (2002), who suggested an alternative 
explanation that the FeH residing below the cloud deck can be seen through 
holes in the clouds.  
  
The line spectra depend not only on the stratification of the 
molecule of interest but also on the nature of the background opacities 
including  dust, H$_2$ CIA, resonance wings of K I and Na I etc in addition
to the usual continuous opacities. In the photospheres of ultracool dwarfs,
all these quantities show drastic changes with $T_{\rm eff}$, log $g$, 
wavelength region etc., and their effect upon the spectral lines can best 
be investigated by the spectral line intensities outlined above. 
Of course, all these effects are automatically taken into account in
the computation of the synthetic spectra (Sect.\,3.2), but the dependence
of the spectral features on various physical parameters can be more clearly
realized in the simple line intensities. Also, the spectral line intensities 
can directly be used for abundance analysis if sufficiently weak lines can 
be measured at high resolution, or can be used as the abscissa of the 
curves-of-growth for the general cases.

\subsection{ Synthetic Spectra and Spectral Energy Distributions}

In the computation of the spectra, we  apply the linelist 
including H$_{2}$O, $^{12}$CO, $^{13}$CO, OH, SiO, CN, and K\,I while 
other molecules including CH$_4$, NH$_3$, PH$_3$, H$_2$S, CO$_2$, TiO, 
and VO are treated as  pseudo-continua with the use of the band model 
method. Now, in applying these spectra to an analysis of the actual 
spectra, we try some improvements over the previous work (Paper I):
First, we now use the linelist of H$_2$O based on the work of 
\citet[][]{par97} instead of the HITEMP database \citep[][]{rot97} used before.
The resulting spectra, however, show rather minor change in the spectral 
region and at the resolution we are interested in. Second, we  change the 
$f_e$-value of FeH from 0.013 \citep[][]{lan90} to an empirical value of 
0.001 (Schiavon, Barbuy, \& Singh 1997), and the resulting spectra turned 
out to show 
better agreement with the observed ones. Also, we replaced the band model 
opacity with the linelist by \citet[][]{phi87} and with the intensity data 
by \citet[][]{sch97}, which are both made available by Samner Davis.

Third, the most serious problem is
that CH$_4$ is quite dominant especially in T dwarfs as well as in late
L dwarfs, but the band model opacity largely overestimates the
methane absorption except for the latest T dwarfs. We then tried the
linelist of CH$_4$ included in the GEISA database \citep[][]{jac99}
but it generally underestimates the methane absorption, since it is 
mainly for application to the Earth's atmosphere of $T \approx 300$\,K.
We have no final solution for methane opacity at present and
we apply the two alternative methods: one using the band model opacity of 
CH$_4$ as well as of FeH and the other the linelists for CH$_4$ as well as
FeH, which we refer to as as case I and case II, respectively.
The resulting predicted spectra or spectral energy distributions (SEDs)
are discussed in Sect.\,4 in comparison with the observed spectra.

\subsection{Infrared Colors}
 With the synthetic spectra or SEDs discussed in Sect.\,3.2, integrated
flux over a filter band can be evaluated by applying an appropriate 
filter response function. As an example, we apply the filter response function
$S_{\rm band}(\lambda)$ of the MKO system \citep[][]{tok02,sim02} to
$F(\lambda)$ based on the UCMs ($T_{\rm cr} = 1800$\,K) and evaluate
$$ F_{\rm band} = \int_{\lambda_1}^{\lambda_2} S_{\rm band}(\lambda)
                    F(\lambda)d\lambda, \eqno(15) 
$$  
where ${\lambda_1}$ and ${\lambda_2}$ are the lower and upper limits,
respectively, of the response function. We apply the case I methane opacity, 
which better reproduces the  CH$_4$ bands where they give the most serious 
effect on $F(\lambda)$, i.e. in T-dwarfs (Sect.\,4).
 
The resulting integrated band fluxes  $F_J, F_H,$ and $F_K$ (in unit of erg 
s$^{-1}$ cm$^{-2}$)  are given in Table 3 on logarithmic
scale, which can be applied to estimate the infrared colors and infrared 
bolometric corrections. For example, $J-K$ can be given by
$$(J-K)_{\rm MKO} = -2.5({\rm log}\,F_{J} - {\rm log}\,F_{K}) + C, \eqno(16)
 $$ 
where the constant is determined to be $C = 1.328 $ so that $(J-K)_{\rm MKO} 
= 0.0$ for Vega by the use of the integrated band 
fluxes for the model of $T_{\rm eff} = 9550$\,K  and log\,$g$ = 3.95
\citep[][]{kur93}, also given in Table 3. The resulting values of the $J-K$ 
index for the three log\,$g$ values are shown in Fig.5, which is an updated 
version of Fig.\,8b (Paper I) for the log $g$ = 5.0 models. Inspection of 
Fig.\,5 reveals that $J-K$ is redder for the higher gravities in the L dwarf 
regime and this can be understood by the gravity effect of the dust mass 
column densities discussed in Sect.\,2.3 (Fig.3). Also, the red limit of $J-K$ 
is almost independent of the gravity and thus our estimation of the value
of $T_{\rm cr} $ based on the red limit of $J-K$ can now be deemed as well 
confirmed for the possible range of gravities of L and T dwarfs.

\section{OBSERVED VS. PREDICTED SPECTRA OF L AND T DWARFS}

The computation of the spectra is done with the step of 0.1\,cm$^{-1}$
for the spectral interval between 0.8 and 2.6\,$\mu$m, and the resulting 
spectra are convolved with the slit function which is assumed to be the 
Gaussian with FWHM = 500\,km s$^{-1}$.  
We assumed $T_{\rm cr} = 1800$\,K throughout.
We have tried other values of $T_{\rm cr} $ but no improved fit could
be obtained in general, and we do not think that it is useful to 
fine tune  such a parameter case by case at present.

\subsection{ Middle L Dwarfs}

We have no sample of the early L dwarfs and we start with the middle L dwarfs
of  L3 - L6.5. Since methane may not be prominent in these L dwarfs, 
we are not bothered by the poorly known data on methane.  
However, we  apply the case II opacities (use linelist of CH$_4$ and FeH), 
since CH$_4$ is already abundant in the models with $T_{\rm eff}$ as high as
1800K (Fig.2a) and methane bands will appear if the methane opacity is 
overestimated by the case I opacities (use the band model opacity for CH$_4$ 
and FeH). 

As an example, the observed spectrum of the L6.5 dwarf 2MASS\,1711+22 shown 
by the filled circles  is 
compared with the predicted ones based on the UCMs of five different sets 
of $T_{\rm eff}$ and log\,$g$ in Fig.\,6. The best fit is obtained for 
$T_{\rm eff} = 1800$\,K and log\,$g = 5.0$  shown by the solid line in 
Fig.\,6c. For comparison, the predicted spectrum based on our model of case C, 
in which dust clouds are effectively cleared up, is shown for the same 
$T_{\rm eff}$ and log\,$g$ by the dashed line.   The difference of the solid 
line against the dashed line shows the effect of thin dust clouds formed in the
layer of $T_{\rm cr} \la T \la T_{\rm cond}$ in the UCM. Clearly the effect 
of the dust extinction is the largest in the $J$ band region and is still 
appreciable in the $H$ band region compared with that in the $K$ band region. 
Thus the effect of the dust clouds is appreciable even though the dust clouds
are rather thin at $T_{\rm eff} \approx 1800$\,K (Fig.\,2a). 
 
The effects of changing 
$T_{\rm eff}$ by $\pm$100\,K at the same log $g$ are shown in Figs.\,6b and d,
and the fits are worse at the $J$ band (note that we first matched the 
observed and predicted spectra at the $K$ band in general). 
The strengths of molecular bands including the  1.4 and 1.9\,$\mu$m H$_2$O 
bands as well as the CO first overtones at 2.3\,$\mu$m can be reasonably
well reproduced by all the UCMs of log $g = 5.0$ shown in Fig.\,6.

The effect of changing log $g$ by +0.5 at $T_{\rm eff} = 1900$\,K is shown 
in Fig.\,6a and a reasonable fit of the overall SED is recovered. However,
the molecular bands turn out to be weaker than in the case of log\,$g = 5.0$
at the same $T_{\rm eff}$ (Fig.\,6b). Also, the effect of  
changing log $g$ by -0.5 at $T_{\rm eff} = 1700$\,K is shown in Fig.\,6e;
the fit of SED is again recovered but the molecular bands appear to be 
strengthened compared with the case of $T_{\rm eff} = 1700$\,K and log\,$g = 
5.0$ (Fig.\,6d). Thus the molecular bands are weaker at the higher gravities, 
which is consistent with the results outlined in Sect.\,3.1 (Fig.\,4), 
and SED shows larger extinction at the higher gravities as can be understood 
by the gravity effect on the dust column densities  noted in Sect.\,2.3 
(Fig.\,3). 

Although the overall  fit of the observed data of 2MASS\,1711, viewed both 
as SED and as spectrum,  can be obtained for the predicted spectrum based 
on the UCM with $T_{\rm eff} = 1800$\,K and log $g = 5.0$, a noticeable gap 
is found at the peak of the $H$ band; the observed spectrum is rather flat 
with some absorption features while the predicted spectra show a smooth 
convex feature.   A possible contribution of the FeH 
$E^4\Pi - A^4\Pi$ system to the absorption features in the $H$ band region
is known \citep[][]{wal01, cus03, nak04}, but we could not include this
system in our predicted spectra because of the lack of the necessary
spectroscopic data. Also, it is not sure if all the absorption features
can be explained by FeH and it is quite possible that other unknown sources
will be important.            

A more or less similar analysis is done for the spectra of 2MASS\,1146+22 
(L3), 2MASS\,1507-16 (L5), SDSS\,2249+00 (L5), and 2MASS\,0920+35 (L6.5),
which were also observed with the Subaru (Nakajima, Tsuji, \& Yanagisawa 
2001). The results are shown in Fig.\,7 in which
the meanings of the solid and dotted lines are the same as in Fig.\,6c. The 
observed spectrum of 2MASS\,1146 (L3V) can be fitted slightly better with the
predicted one for $ T_{\rm eff} = 1900$\,K and log\,$g =5.5$ rather than
that for $ T_{\rm eff} = 1850$\,K and log\,$g =5.0$. This is consistent with 
a possibility that the early L dwarf such as 2MASS\,1146 is a main-sequence 
star rather than a brown 
dwarf as is suggested elsewhere \citep[][]{nak04}.  Also, we show the case
of  $ T_{\rm eff} = 1850$\,K and log\,$g =5.0$ for 2MASS\,1507 (L5) rather
than the case of $ T_{\rm eff} = 1900$\,K and log\,$g =5.5$.
The short segment of the spectrum of SDSS\,2249 (L5) can roughly be fitted 
with the predicted one based on the UCM with $ T_{\rm eff} = 1800$\,K and 
log\,$g =5.0$, and a slightly lower gravity may improve the fit. An
interesting feature is that the $H$ band region can be accounted for rather 
well by the UCM for this object, while it is more difficult to account
for the $H$ band spectra of other objects as noted already. 

In conclusion, the basic features of these L dwarfs shown in Fig.\,7  are 
rather similar to those of 2MASS\,1711 (L6.5) shown in Fig.\,6, and the 
overall SED as well as the strengths of the most molecular bands can be 
fitted with the UCMs of $T_{\rm eff} \approx 1800 - 1900$\,K. One unsolved 
problem in these fits is the gap at the $H$ band region except for SDSS\,2249.

\subsection{Late L Dwarfs}

  The detailed comparison of the observed spectra of L dwarfs revealed
that the H$_2$O bands at 1.1 and 1.4\,$\mu$m are not necessarily stronger
in the L8 dwarf 2MASS\,1523+30 than in the L5 and L6.5 dwarfs, but the CH$_4$
bands at 2.2\,$\mu$m can be identified in the L8 dwarf \citep[][]{nak04}
as well as the stronger bands at 3.3\,$\mu$m \citep[][]{nol00}.
We found that the overall SED as well as the molecular bands of 2MASS\,1523 
(L8) can be fitted reasonably well with the predicted spectrum based on the
UCM of $ T_{\rm eff} \approx 1500$\,K and log\,$g \approx 5.0$ as shown in 
Fig.\,8. The water bands based on the UCMs are rather weak possibly because 
of the large dust extinction. In fact, the dust column density in the
observable photosphere is the largest at about $ T_{\rm eff} \approx 1500$\,K
(Fig.\,3) and  this fact  results in a very large difference of the
emergent spectra based on the UCM (solid lines) and those based on the 
dust-segregated models of case C (dashed lines). Because of this large
dust extinction, H$_2$O 1.1\,$\mu$m bands as well as K\,I 1.2\,$\mu$m
doublet are rather weak in the L8 dwarf 2MASS\,1523.

We have applied two different opacities for the methane bands
(Sect.\,3.2); the 
band model opacity (case I) and the line-by-line opacity (case II), which may 
provide the maximum and minimum estimates of the real opacity, respectively. 
The predicted spectra  show bifurcation in the 1.6 and 2.2\,$\mu$m regions 
and the observed spectrum should appear between the high and low estimates
of the emergent spectra. This expectation is met in the 2.2\,$\mu$m 
bands, but it is clear that the band model opacity (case I) highly
overestimated the methane bands. On the other hand, the predicted spectra
based on the available linelist of CH$_4$ (case II) provides a reasonable fit 
to the observed 2.2\,$\mu$m CH$_4$ bands. The predicted $H$ fluxes are
slightly higher than the observed and it is possible that the unknown 
opacity prevailing in the $H$ band region of L3 - L6.5 dwarfs noted 
in Sect.\,4.1 may have some effect at L8 as well.
It is to be noted, however, that the 1.63 and 1.67 $\mu$m absorption 
features can be seen both in the observed and predicted (case II) spectra.

\subsection{Early T dwarfs or L/T Transition Objects }

SDSS\,1254-01 is one of the three remarkable objects whose infrared colors 
and spectra are both intermediate between the late L dwarfs and cool T dwarfs
\citep[][]{leg00}, and SDSS\,1254 is now classified as T2 \citep[][]{bur02a}. 
The predicted spectra based on the UCMs with $T_{\rm eff}= 1400, 1300$, 
and 1200\,K are shown in Fig.\,9, and the effect of dust extinction decreases 
in this order as evidenced by the decreasing difference between the solid 
and dashed lines whose meanings are as noted already. On the other hand, 
the predicted intensities of the molecular bands increase according as 
the $T_{\rm eff}$ decreases.
The best fits, both in the overall SED and in the molecular band strengths,
are found for the case of the UCM with  $ T_{\rm eff} \approx 1300$\,K and 
log $g \approx 5.0$. The observed methane bands, both  at 1.6 and 2.2\,$\mu$m,
are just between the predictions based on the cases I and II opacities.

In our UCMs, the dust cloud is partly immersing into the optically thick 
region in the early T dwarfs, and this is due to  a natural consequence 
that the dust condensation temperature is just near the optical depth 
unity at about this spectral type. As a result, the effect of dust 
extinction is not so large as in the late L dwarfs while volatile 
molecules including  CH$_4$ can be formed in the layer above the dust clouds.  
Also, the position of SDSS\,1254 on the CM diagram could be reasonably
well reproduced by our UCMs with the same  value of $ T_{\rm cr} = 
1800$\,K \citep[][]{tsu03a}. Thus it should be emphasized that the very
simple assumption in our UCMs that the segregation of the dust grains 
takes place at about $T_{\rm cr} \approx 1800$\,K throughout L and T
dwarfs accounts for the rapid bluing in the L/T transition as well as the 
rather unique spectra of the transition object.   

\subsection{Middle T dwarfs  }

The observed spectrum of the T3.5 dwarf SDSS\,1750+17 is compared with the 
predicted ones of  $ T_{\rm eff} =  1200, 1100$ and 1000\,K in Fig.\,10.  
The overall SED appears to be fitted reasonably well with the UCM of  
$ T_{\rm eff} \approx 1100$\,K and log\,$g \approx 5.5$, and the higher
gravity is preferred to explain the rather weak water bands. The 
predicted water bands  still appear to be stronger than the observed ones,
and it is possible that the chemical composition of this object may be
non-solar. The observed methane bands are well between  cases I and II, but
the 1.6\,$\mu$m bands are closer to case I while the 2.2\,$\mu$m bands
to case II. At these low effective temperatures, the effect of dust clouds 
is rather minor as can be seen in that the SEDs for the cases with
and without cloud, shown by the solid and dashed lines, respectively,
do not differ significantly. Also, the overall SED can be fitted with the 
cloud-free model of $ T_{\rm eff} =  1200$\,K (dashed line in Fig.\,10a).
It is difficult to judge which of these different cases provides the better 
fit to the observation.  However, the presence of the dust cloud is
indispensable for L and early T dwarfs, and we think that it is reasonable to
assume the same models for the later T dwarfs as well.
 
The observed spectrum of the T3.5 dwarf SDSS\,1750 may look rather 
similar to that of SDSS\,1254 except that the methane bands are stronger.  
The overall SED of SDSS\,1750, however, is definitively different from
those of the L and L/T transition objects and already shows the typical 
characteristic of T dwarfs in that the flux peaks (in $f_{\nu}$ unit) at $J, 
H,$ and $K$ bands decrease in this order, namely $f_{J} > f_{H} > f_{K}$
(also see Fig.\,11). For comparison,  L dwarfs show just the opposite in 
that $f_{J} < f_{H} < f_{K}$ (see Figs.\,6 - 8), and the L/T transition object 
SDSS\,1254 shows the intermediate behavior of L and T dwarfs in that 
the $H$ flux is the highest, namely  $f_{J} < f_{H} > f_{K}$ (see Fig.\,9). 
These gross features are well reproduced by our UCMs (solid lines throughout
Figs.\,6-11 with the case I methane opacity to the 1.6\,$\mu$m methane 
bands in T dwarfs). It is to be noted that the SEDs show $f_{J} > f_{H} > 
f_{K}$ throughout L and T dwarfs if there is no cloud (i.e. our case C shown 
by the dashed lines throughout Figs.\,6 - 11) because of the large infrared
opacity due to H$_2$ CIA in ultracool dwarfs. The depressions of the $J$ flux 
in the L/T transition objects and further of the $H$ flux in the L dwarfs
are due to the rise of the dust clouds to the optically thin region, which
is a natural consequence of the basic assumption of UCM that the dust clouds 
form only in the region of $ T_{\rm cr} \la  T \la  T_{\rm cond}$.

\subsection{Late T dwarfs}
 
We include the classical T dwarf Gl\,229B (T6), the observed spectrum 
\citep[][]{geb96} of which is found to show a reasonable fit to the 
predicted one based on the UCM with $ T_{\rm eff} = 900$\,K and log $g = 
5.0$ as shown in Fig.11a. The observed methane bands at 2.2\,$\mu$m are 
still between the predicted ones based on the cases I and II methane 
opacities, but the 1.6\,$\mu$m bands can be fitted with the predicted 
spectrum based on the band model opacity (i.e. case I).
In a recent paper, \citet[][]{burr02} showed that the spectrum of
Gl\,229B  in the short wavelength region near 1\,$\mu$m could be fitted with
their models of  $ T_{\rm eff} = 950$\,K/log\,$g = 5.5$ as well as of
$ T_{\rm eff} = 750$\,K/log\,$g = 5.0$. We could fit the same 
region with our model of $ T_{\rm eff} = 900$\,K/log\,$g = 5.0$, and 
slightly different results may be partly because we are still
using the classical Lorentzian profiles for the K\,I opacity,
while a more sophisticated theory of the line 
broadening must be called for \citep[][]{burr00,burr03}.

For the latest T dwarf in our sample, 2MASS\,1217-03 (T7.5), 
the methane bands, both  at 1.6 and 2.2\,$\mu$m, can roughly be 
accounted for by the band model opacity (case I) rather than
the line-by-line opacity (case II) as shown in Fig.\,11b.
The spectrum of 2MASS\,1217 shows stronger bands
of methane as well as of water than in Gl\,229B, and can be fitted with 
the predicted spectrum based on the lower $ T_{\rm eff}$ of 800\,K and 
the lower gravity of log\,$g = 4.5$, as in Fig.\,11b. Also, the same
spectrum can marginally be fitted with 
the predicted one based on the higher $ T_{\rm eff}$ of 900\,K and 
the higher gravity of log $g = 5.0$. It is interesting that  the
same observed spectrum can be fitter either by higher $ T_{\rm eff}$/
higher gravity or by lower $ T_{\rm eff}$/lower gravity (see also 
Fig.6), and the same effect was shown by \citet[][]{burr02}
as noted above for Gl\,229B.   Thus accurate  estimation of gravity
from the infrared spectrum  may be difficult unless  $ T_{\rm eff}$
can be determined by other methods. 

In the model of $ T_{\rm eff} \la 1000$\,K, the predicted spectrum based
on our UCMs differs little from that of the fully dust-segregated
model of case C as noted in Sect.\,4.4. Thus, dust clouds give almost
no effect on the observed spectrum once the immersion of the dust 
clouds in the optically thick regime is  completed in these very cool 
models. This result is consistent with the earlier observations
by \citet[][]{lie00} who showed that dust gives little effect on the
observed spectrum  of the late T dwarf SDSS\,\,1624+00. Thus our 
previous proposition that the warm dust, together with the
the K\,I and Na\,I resonance lines, may produce observable
effect on the spectra of cool T dwarfs such as Gl\,229B \citep{tsu99}
cannot be supported, even if the warm dust clouds exist in the deeper layer. 
This result implies that observational studies of dust in cool T dwarfs
should be quite difficult.

\section{DISCUSSION}

\subsection{Modeling}

The UCM used in this paper is a kind of semi-empirical models
rather than a fully consistent theoretical model. This approach is
based on the recognition that it should be more difficult to treat
all the processes taking place in the dusty photosphere in which
phase changes in gas, liquid, and solid may induce complicated
chaotic phenomena. An extreme case is the Earth's atmosphere
which embraces all the complicated phenomena treated by another
big field of science - meteorology.  Instead of pursuing the
detailed microscopic processes of dust formation and destruction,
we tried to approximate the resulting possible structure of the
cloudy photosphere by a model to be treated within the framework of the
classical non-grey theory. For this purpose, we introduced a
simple assumption that the dust grains formed at its condensation
temperature will soon grow too large to be sustained in the photosphere
at a slightly lower temperature which we referred to as the critical 
temperature. The only parameter introduced in our semi-empirical approach 
is the critical temperature, in addition to the mixing length which is 
assumed to be one pressure scale height in treating convection. 

We should certainly do our best to minimize the number of free-adjustable 
parameters in such a semi-empirical approach, since any
observed data may be ``explained'' if many parameters are assumed. In this  
paper, we tried to see to what extent the UCMs with the empirically fixed
unique  value of $T_{\rm cr} = 1800$\,K throughout can explain the
available observed spectra.  It was not expected from the beginning that
the fits can be perfect for such a simplified treatment, and further
because of the many approximations both in the model itself as well as 
in the input data (Sect.\,5.2). Nevertheless the overall characteristics 
of the SEDs as well as the major spectroscopic features of L and T  
dwarfs  can be reasonably accounted for (Figs.6-11).  Also, infrared 
colors (Paper I), L/T transition \citep[][]{tsu03a}, and L-T spectral 
classification \citep[][]{tsu03b, nak04} can reasonably be interpreted with 
the UCMs. Thus, the basic assumption of the UCMs can be deemed as 
well supported by the observations.   

What is important to conclude from the reasonable agreement between
the major observational data of L-T dwarfs and predictions from the UCMs 
is that the dust should certainly form in the photospheres of cool dwarfs 
but only a small amount of dust should be sufficient. In fact, if dust forms
in the full amount as predicted by the thermochemistry, the photosphere will
soon be filled in by dust in the cooler brown dwarfs and its spectrum will 
look like a blackbody as in our case B models (e.g. Paper I). 
Also this small amount of dust should be concentrated rather deep in
the photosphere, since  its effect should appear in the coolest T dwarfs if 
a small amount of dust is distributed uniformly throughout the photosphere. 
For this reason, we assumed that the dust should be in the form of a thin
cloud deep in the photosphere. The idea that a finite-thickness cloud
could explain the approximate shape of the $M_J$ vs. $J-K$ diagram was
also proposed by \citet[][]{mar00}  based on a different approach. 
Thus the next problem in modeling is to understand why the photospheres of 
cool dwarfs adjust themselves in such a way as to produce only a small 
amount of dust in a form of the thin cloud.  

As a possible mechanism to produce the cloud, convection may play a role as 
discussed by several authors \citep[e.g.][]{ack01, hel01, mar02, coo03}. 
One interesting feature is that the particle sizes can be determined
by considering the time scale of the convective dredge-up of the 
raw material to the dust forming region. However,
the convective zone is situated rather deep in the photosphere
\citep[e.g.][]{burr97, all01, tsu02}, and it is not necessarily  
possible that the convection will reach so nicely to the dust forming 
region in all the models of L and T dwarfs. The present convective models 
are based on the mixing-length theory (MLT), but recent detailed 2D and 3D
hydrodynamical simulations of surface convection in a late M-dwarf
\citep[][]{lud02} confirmed that the classical
MLT allows reasonably accurate prediction of the thermal structure
of the late M dwarf and that overshooting extends the convective mixing
region only modestly (about 2 pressure scale heights) beyond the 
Schwarzschild boundary. The possible interplay between convection and cloud 
formation may be an interesting subject to be pursued further, but
our assumption in UCMs is that the thermodynamical constraint, as the
first approximation, determines the basic feature of the dusty photospheres.

\subsection{Input Data}

Apart from the fundamental problem in modeling, the input data are
still far from satisfactory. One serious problem is the methane opacity. 
Although it appeared that the presently available linelist roughly accounts 
for the observed intensities of the 2.2\,$\mu$m bands throughout late L to 
middle T dwarfs (Figs.\,8-10), this may be only fortuitous. 
In fact, the details of the predicted spectra based on the present linelist
\citep[][]{jac99} can never be fitted well with the observed spectra, as
shown in Fig.\,12 for SDSS\,1750 as an example. Inspection of Fig.\,12
reveals that only a limited number of predicted bands show 
correspondences with the observed ones, and it is clear that many bands 
are missing in the present linelist. Our previous conclusion that
the band model opacity may be preferred (Paper I) is only applicable
to late T dwarfs in which methane bands are quite strong, as is confirmed
in the coolest T dwarf 2MASS\,1217 in our sample (Fig.\,11b),
but cannot be justified for most cases as are evident in Figs.\,6 - 10.  

While the problem of molecular opacities can be solved mostly if
a more complete linelist can be provided either by experiments or
by theories, the case of the dust opacities may be more difficult,
since it is closely related to the cloud formation itself. 
For example, the chemical equilibrium abundance pattern of the dust grains
in the stratified clouds suffers the effect of depletion \citep[][]{lod99}
and  the so-called rainout \citep[][]{burr99}. Also, the equilibrium gas and 
dust chemistry at low temperatures is quite complicated and involves many 
problems that require detailed analyses \citep[e.g.][]{lod02, lod02b}. 
Also, possible effects such as due to impurities (the so-called dirty grains) 
and the core-mantle structures may introduce further difficulties.  
It is not possible to incorporate all these complications in the present 
modeling and we restricted ourselves to consider only a few most abundant 
condensates as noted in Paper I. Probably, it should be required to consider 
the dust opacities more carefully to have a better fit between observed and
predicted SEDs. Unfortunately, it is more difficult to improve the
situation because dust, unlike atoms and molecules, shows few direct 
spectroscopic features and thus it is very difficult to have empirical 
assessments on the dust opacities.  

\subsection{Applications}
   
The effective temperatures corresponding to the best fits between the 
observed and predicted spectra discussed in Sect.\,4 are summarized in  
Table 4 as $T_{\rm eff}$(SED). Although the resulting values of  $T_{\rm eff}$
show little change within the middle L dwarfs (L3 - L6.5), they show steady 
decrease to the late L and further to the early, middle and late T dwarfs. 
Thus the observed characteristics of the SEDs and spectra are reasonably 
interpreted as the temperature effect by the UCMs. The resulting 
$T_{\rm eff}$ values  are also  compared in Table 4 with those obtained from
the bolometric fluxes based on the integrated near infrared fluxes, 
$T_{\rm eff}(f_{\rm bol})$, and on the $K$ band bolomtric correction, 
$T_{\rm eff}({\rm BC}_K)$, in the separate paper 
\citep[][]{nak04}, which applied the same models to obtain the bolometric 
corrections but emphasized  the different aspects of both the observed and 
predicted data.  The agreement of the $T_{\rm eff}$ values based on the
different methods is generally fair except for L5 dwarf 2MASS\,1507, and 
this fact may confirm the mutual consistency of our analyses.

As for the L5 dwarf 2MASS\,1507, the present spectral analysis shows 
$T_{\rm eff} \approx 1850$\,K while the result based on the bolometric 
flux  shows $T_{\rm eff}$ value as low as 1400\,K. One problem is that 
the same infrared color does not necessarily correspond to the same 
effective temperature (e.g. Fig.\,5) or to the same spectral type 
\citep[e.g.][]{kir00,leg02}, and this fact implies that the similar SED 
may result from the different values of $T_{\rm eff}$ or different spectral 
types. In other words, our result based on the SED is not necessarily a 
unique solution but there may be a different solution closer to 1400\,K. 

To examine such a possibility, we compare the observed spectrum of 2MASS\,1507
with the predicted ones from the high temperature models ($T_{\rm eff} = 
1700 - 1900$\,K) in Fig.13a and with those of the low temperature models 
($T_{\rm eff} = 1400 - 1600$\,K) in 
Fig.13b. Inspection of Fig.13a suggests that the overall shape of the SEDs 
as well as the major molecular bands such as of CO and
H$_2$O can reasonably be accounted for by a model with $T_{\rm eff}$ between
1800 and 1900\,K and, for this reason, we suggested $T_{\rm eff} \approx 
$ 1850\,K in Sect.\,4.1 (Fig.\,7b). On the other hand, the overall SED
appears to be accounted for by a model of $T_{\rm eff} \approx $ 1400\,K 
in Fig.13b. However, it also appears that the methane bands at 2.2\,$\mu$m
is predicted to be quite appreciable by this model. This computation of
methane bands is based on the linelist of CH$_4$ (case II) and not due to the
overestimation by the band model opacity referred to as case I (see Fig.\,9a 
for the predicted spectrum based on the case I opacity for the 1400\,K model). 
For this reason, we 
cannot accept the low temperature model for the L5 dwarf 2MASS\,1507, and
the origin of the discrepant $T_{\rm eff}$ values by the different methods
remains unsolved.  In conclusion, even though the same infrared
color corresponds two different effective temperatures, this degeneracy
can be removed in the spectra by considering both the overall shape of the
spectrum (or SED) and some molecular features sensitive to temperature (e.g.
CH$_4$).

Also our UCMs were used to interpret the CM diagram such as ($J-K, 
M_J$) diagram\citep[][]{tsu03a}. For this attempt, there is a severe 
criticism  that the detailed 
behavior of the models on the CM diagram does not match observations 
\citep[e.g.][]{tin03}. However, the point of our present analysis based on 
UCMs is not the detailed quantitative fits to individual objects, but rather
directed to understand the overall behaviors of the colors, magnitudes, SEDs,
and spectra throughout L to T dwarfs based on a single sequence of model 
photospheres. For this purpose,  
our results  provided a possibility of unified understanding
of all these observables while the previous models (our models B and C 
as well as more or less similar models by other authors) could not.
As for ($J-K, M_J$) diagram, the $J$ band flux suffers the most serious 
effect of dust and the difficulties such as noted in Sects.5.1 and 5.2 must 
be overcome before we can achieve a better quantitative fit. 

Finally,  appropriate knowledge on the $p-T$ structure of the photospheres
should be vital to analyze high resolution spectra of L and T dwarfs.
So far, we have restricted ourselves to examine the effect of $T_{\rm eff}$ 
and log\,$g$ on the spectra, but abundance should certainly be another factor 
to be considered. For example, observed H$_2$O bands in SDSS\,1750 appeared to
be too weak to be explained by the predicted ones (Sect.\,4.4), unlike the
other objects, and such a possibility that the oxygen abundance and/or 
metallicity in SDSS\,1750 may be non-solar can be confirmed only by the
detailed abundance analyses.  Since the line broadenings by 
turbulence, damping, and other effects must be considered simultaneously
for this purpose, quantitative analysis of high resolution spectra 
should be called for. It is to be noted that the recent progress in the IR 
spectroscopy finally made it possible to analyze high resolution infrared 
spectra of faint brown dwarfs \citep[e.g.][]{smi03}.

\section{CONCLUDING REMARKS}

We have shown that the spectra or SEDs of L and T dwarfs
can be interpreted consistently by a single grid of UCMs.
At present, we cannot yet achieve a fully self-consistent
model photosphere of ultracool dwarfs because of the complexities 
due to the coupling of physico-chemical processes relating to the 
cloud formation and associated dynamical processes. Instead, we
restricted ourselves to a semi-empirical approach which is based only on
a simple thermodynamical constraint, and reduced all the possible
complicated dynamical effects to a quasi-static model photosphere
to be treated by the classical non-grey theory. 
 It is to be noted that the model photosphere itself is not necessarily our
final purpose, but our purpose is to understand the real astronomical
objects, in this case, L and T dwarfs. The model photosphere is simply a 
means by which to help this aim, even though better models are certainly
more useful for this purpose. Thus the aim of our UCMs is not to
provide the exact quantitative fits to observed data at present.
It is hoped that our semi-empirical
approach can be of some help as a guide to interpret and analyze the 
observed data of ultracool dwarfs, and hopefully will provide a guideline 
by which a more physical model can be developed in the near future.
To be of some use for this purpose, the numerical data of the UCMs, 
including the spectra and SEDs, are  made available through our Web site
\footnote{http://www.mtk.ioa.s.u-tokyo.ac.jp/\~\,ttsuji/export/ucm}.

\acknowledgements
One of the authors (T.T.) would like to thank Sumner Davis for allowing
to access his spectroscopic database of molecules
including FeH and Hugh Jones for his help on the spectroscopic database
of H$_2$O lines by Partridge \& Schwenke. We thank the Subaru staff and
all those who have contributed to develop the fine spectrographs
as well as the Subaru telescope, with which our observations could
have been made possible. We also thank an anonymous referee for helpful
comments and suggestions. This work was  supported by the grants-in-aid of 
JSPS Nos.11640227 (T.T.) and 14520232 (T.N.).

\clearpage

\begin{table}
\caption{PHYSICL PARAMETERS OF DUST GRAINS} 
\begin{tabular}{lcccccc}
\noalign{\bigskip}
\tableline\tableline
\noalign{\bigskip}
 dust species  &  $a_{0}\,({\rm \AA})$\tablenotemark{a} &  
$\sigma$\,(dyn\,cm$^{-1})$\tablenotemark{a} &
$\rho_{\rm dust}$\,(gr\,cm$^{-3})$\tablenotemark{b}  & 
 $n^{*}$\tablenotemark{c}  &  $r_{\rm cr}\,({\rm \AA})$\tablenotemark{c} 
&~~$Q_{\rm ext}~ \&~ \gamma $\tablenotemark{d}\\
\noalign{\bigskip}
\tableline
\noalign{\bigskip}
corundum (Al$_2$O$_3)$  & 1.7179  &  690 &  4.022 & 3.2$\times 10^{5}$ & 118 
& 1  \\
iron (Fe)           & 1.4114  & 1800 &  7.874 & 1.7$\times 10^{6}$ & 171 
& 2, 3, 4 \\
enstatite (MgSiO$_3$)  & 2.3193  &  400 &  3.209 & 3.8$\times 10^{5}$ & 168 
& 5  \\
\noalign{\bigskip}
\tableline
\end{tabular}

\tablenotetext{a}{Hasegawa \& Kozasa (1988)}
\tablenotetext{b}{Weast (1985-86)}
\tablenotetext{c}{for the supersaturation ratio  $S = 1.1$}
\tablenotetext{d}{Numerical results (Table 2) are based on the optical 
constants by:
 (1) Eriksson et al.\,1981;
 (2) Lenham \& Treherne\,1966;
 (3) Johnson \& Christy\,1974;
 (4) Ordal et al.\,1988;
 (5) Ossenkopf, Hennings, \& Mathis\,1992. 
}
\end{table}
\clearpage

\begin{table}
\caption{ EFFICIENCY FACTOR FOR EXTINCTION AND ALBEDO ( $ r = 0.01\mu$m) }
\begin{tabular}{ccccccccc}
\noalign{\bigskip}
\tableline\tableline
\noalign{\smallskip}
\noalign{\smallskip}
 $\lambda$ & iron~~~ & & & enstatite & & & corundum & \\
  \cline{2-3}\cline{5-6}\cline{8-9}
  $(\mu$m)  &  log\,$Q_{\rm ext}$ &  $\gamma$ &   &
log\,$Q_{\rm ext}$   &  $\gamma$  &  & log\,$Q_{\rm ext}$ &  $\gamma$ \\
\noalign{\smallskip}
\noalign{\smallskip}
\tableline
\noalign{\bigskip}
  0.100 &    0.520  &   0.251 & &   -0.385  &   0.130 & &   -1.140  &   1.000\\
  0.300 &   -0.341  &   0.090 & &   -1.192  &   0.051 & &   -2.738  &   0.795\\
  0.500 &   -1.095  &   0.008 & &   -1.821  &   0.009 & &   -3.649  &   0.354\\
  0.700 &   -1.334   &  0.004 & &   -2.080  &   0.004 & &   -3.733   &  0.120\\
  0.900 &   -1.545  &   0.002 & &   -2.224  &   0.002 & &   -3.788  &   0.061\\
  1.100 &   -1.707   &  0.002 & &   -2.290 &    0.002 & &   -3.805  &   0.022\\
  1.300 &   -1.836  &   0.002 & &   -2.307 &    0.001 & &   -3.807  &   0.012\\
  1.500 &   -1.964  &   0.002 & &   -2.339  &   0.000 & &   -3.825  &   0.006\\
  1.700 &   -2.092  &   0.001 & &  -2.400   &  0.000  & &  -3.785   &  0.004\\
\noalign{\smallskip}
\noalign{\smallskip}
   2.000 &   -2.285  &   0.001 & &  -2.469  &   0.000 & &  -3.725 &    0.002\\
   3.000 &   -2.765  &   0.000  & & -2.742 &    0.000  & & -3.525 &    0.000\\
   4.000 &   -3.076  &   0.000  & & -2.913 &    0.000  & & -3.439 &    0.000\\
   5.000 &   -3.320  &   0.000 & &  -3.009  &   0.000 & &  -3.460  &   0.000\\
   6.000 &   -3.528  &   0.000 & &  -3.045  &   0.000 & &  -3.267  &   0.000\\
   7.000 &   -3.684  &   0.000  & & -3.031  &   0.000  & & -3.173  &   0.000\\
   8.000 &   -3.834  &   0.000  & & -2.901  &   0.000  & & -3.095  &   0.000\\
   9.000 &   -3.924  &   0.000  & & -2.421  &   0.000  & & -3.002  &   0.000\\
  10.000 &   -4.005  &   0.000  & & -2.067  &   0.000  & & -2.464  &   0.000\\
  11.000 &   -4.202  &   0.000  & & -2.219  &   0.000  & & -1.936  &   0.000\\
  12.000 &   -4.399  &   0.000  & & -2.464  &   0.000  & & -1.716  &   0.000\\
  13.000 &   -4.535   &  0.000  & & -2.656 &    0.000 & &  -1.837  &   0.000\\
  14.000 &   -4.612  &   0.000  & & -2.684 &    0.000 & &  -1.987  &   0.000\\
  15.000 &   -4.677  &   0.000  & & -2.683  &   0.000  & & -2.117 &    0.000\\
\noalign{\smallskip}
\noalign{\smallskip}
  20.000 &   -4.934  &   0.000  & & -2.575  &   0.000  & & -2.553 &    0.000\\
  25.000 &   -5.097  &   0.000 & &  -2.801  &   0.000  & & -2.690 &    0.000\\
  30.000 &   -5.210  &   0.000 & &  -2.923  &   0.000  & & -2.945 &    0.000\\
  35.000 &   -5.297  &   0.000  & & -3.060 &    0.000 & &  -3.123 &    0.000\\
  40.000 &   -5.369  &   0.000  & & -3.171 &    0.000 & &  -3.236 &    0.000\\
  45.000 &   -5.429  &   0.000 & &  -3.275 &    0.000 & &  -3.349 &    0.000\\
  50.000 &   -5.483  &   0.000 & &  -3.380 &    0.000 & &  -3.462 &    0.000\\
\noalign{\bigskip}
\tableline
\end{tabular}
\end{table}

\clearpage

\begin{table}
\caption{LOGARITHMS OF THE INTEGRATED FLUXES OVER THE FILTER BANDS ( 
MKO SYSTEM) }
\begin{tabular}{cccccrccc}
\noalign{\bigskip}
\tableline\tableline
\noalign{\bigskip}
~log $g$~ &  ${T_{\rm eff}}$\,(K) & log\,$F_J$ & log\,$F_H$  & log\,$F_K$ &
~~~~~~~~~~${T_{\rm eff}}$\,(K)   & log\,$F_J$ & log\,$F_H$  & log\,$F_K$  \\
\noalign{\bigskip}
\tableline
\noalign{\bigskip}
4.5   & 700. &  5.902 &    5.513 &  5.034 &  800.&  6.159 &  5.741 &  5.349  \\
      & 900. &  6.364  & 5.959  & 5.622  & 1000. & 6.542 & 6.171 & 5.853 \\
      & 1100. &  6.675 &  6.406 & 6.072  & 1200. &  6.757  & 6.667 &  6.301  \\
      & 1300. & 6.832 & 6.881 & 6.522  & 1400. &  6.860 &  7.036 & 6.776 \\
      & 1500. &  6.908 &  7.138 &  6.981 & 1600. &  6.919 &  7.209 &  7.152  \\
      & 1700. &  7.130 & 7.354  & 7.252  & 1800. & 7.338  & 7.459  & 7.319  \\
      & 1900. &  7.486 &  7.532 &  7.344 & 2000. &  7.623 &  7.616 &  7.405  \\
      & 2100. &  7.716 &  7.688  & 7.466  & 2200. &  7.782 & 7.764 & 7.547 \\
      & 2300. &  7.840 &  7.833 & 7.617  & 2400. &  7.899 &  7.902 &  7.685 \\
      & 2500. &  7.954 &  7.969 &  7.747 & 2600. &  8.008 &  8.035 &  7.807 \\ 
      &       &        &        &        &       &        &        &   \\
  5.0 & 700.  &  5.895 &  5.632 &  5.016 & 800.  & 6.161  & 5.858  & 5.348  \\
      &  900. & 6.378  & 6.064  & 5.620 & 1000. &  6.555 & 6.266  & 5.859 \\
      & 1100. &  6.683 & 6.485  & 6.089 & 1200. &  6.780 &  6.702 & 6.303  \\
      & 1300. &  6.846 & 6.902 & 6.522 & 1400. & 6.884  & 7.058 & 6.760 \\
     & 1500. &  6.885 &  7.152 & 7.000 & 1600. &  6.923 &  7.228 &  7.165 \\
     & 1700. & 7.022 & 7.330 &  7.278 & 1800. & 7.227 & 7.449 & 7.351 \\
     & 1900. & 7.419 & 7.540 & 7.391 & 2000. & 7.562 & 7.620 & 7.437  \\
     & 2100. & 7.695 & 7.706 &  7.482 & 2200. & 7.791 &  7.780 &  7.537  \\
     & 2300. &  7.865 &  7.848 & 7.596 & 2400. & 7.919 & 7.916 & 7.669 \\
     &  2500.&  7.969 & 7.979 & 7.735 & 2600. & 8.018 & 8.041 & 7.797 \\
      &       &        &        &    &        &       &       &    \\
 5.5 & 700.  & 5.867  & 5.713  & 4.943 & 800.  & 6.146  & 5.951  & 5.328 \\  
     &  900.  & 6.374  & 6.152  & 5.608 & 1000. & 6.558  & 6.343  & 5.849  \\
     & 1100. & 6.699  & 6.540  & 6.074 & 1200. & 6.795  & 6.743  & 6.300  \\
     & 1300. & 6.861  & 6.930  & 6.523 & 1400. &  6.910 & 7.076  & 6.743  \\
     & 1500. & 6.940  & 7.183  & 6.962 & 1600. & 6.974  & 7.260  & 7.139  \\
     & 1700. & 7.038  & 7.346  & 7.274 & 1800. & 7.155  & 7.428  & 7.358  \\ 
     & 1900. &  7.366 & 7.529  & 7.403 & 2000. & 7.536  & 7.617  & 7.437  \\
      & 2100. & 7.666  & 7.699  & 7.479 & 2200. & 7.772  & 7.781  & 7.530 \\
     & 2300. & 7.856  & 7.855  & 7.587 & 2400. & 7.925  & 7.922  & 7.648   \\
    & 2500. & 7.983  & 7.987  & 7.712  & 2600. & 8.030  & 8.048  & 7.779  \\
     &       &        &        &        &       &        &        &   \\
 3.95  & 9550. & 9.330  & 9.182  & 8.799 &      &       &        &    \\
\noalign{\bigskip}
\tableline
\end{tabular}
\end{table}

\begin{table}
\caption{ EFFECTIVE TEMPERATURES}
\begin{tabular}{lllll}
\noalign{\bigskip}
\tableline\tableline
\noalign{\bigskip}
Object & Sp. type~~~ & ${T_{\rm eff}}$(SED)~~~ &
   ${T_{\rm eff}}(f_{\rm Bol})$~~~ &  ${T_{\rm eff}}$(BC$_K$)~~~ \\
\noalign{\bigskip}
\tableline
\noalign{\bigskip}
   2MASS\,1146+22A &L3 & 1900\,K  & 1612-1748\,K & 2098-2276\,K\\ 
   2MASS\,1507-16 & L5 & 1850 &  1371-1487 & 1544-1675 \\
   SDSS\,2249+00 & L5 & 1800 &       &  \\
   2MASS\,1711+22 & L6.5 & 1800 &     & \\
   2MASS\,0920+35 & L6.5 & 1800 &     & \\
  2MASS\,1523+30 & L8 & 1500  & 1287-1395 & 1330-1442\\ 
   SDSS\,1254-01 & T2 & 1300  & 1252-1358 & 1279-1387  \\
               &    &       & 1348-1462 & 1348-1462  \\
  SDSS\,1750+17 & T3.5 & 1100 &          &   \\
   Gl229B &  T6 & ~900  & ~905-981   & ~928-1007 \\
  2MASS\,1217-03 & T7.5 & ~800 & ~885-960  & ~873-947      \\
\noalign{\bigskip}
\tableline
\end{tabular}
\end{table}

\begin{figure}
\epsscale{0.65}
\plotone{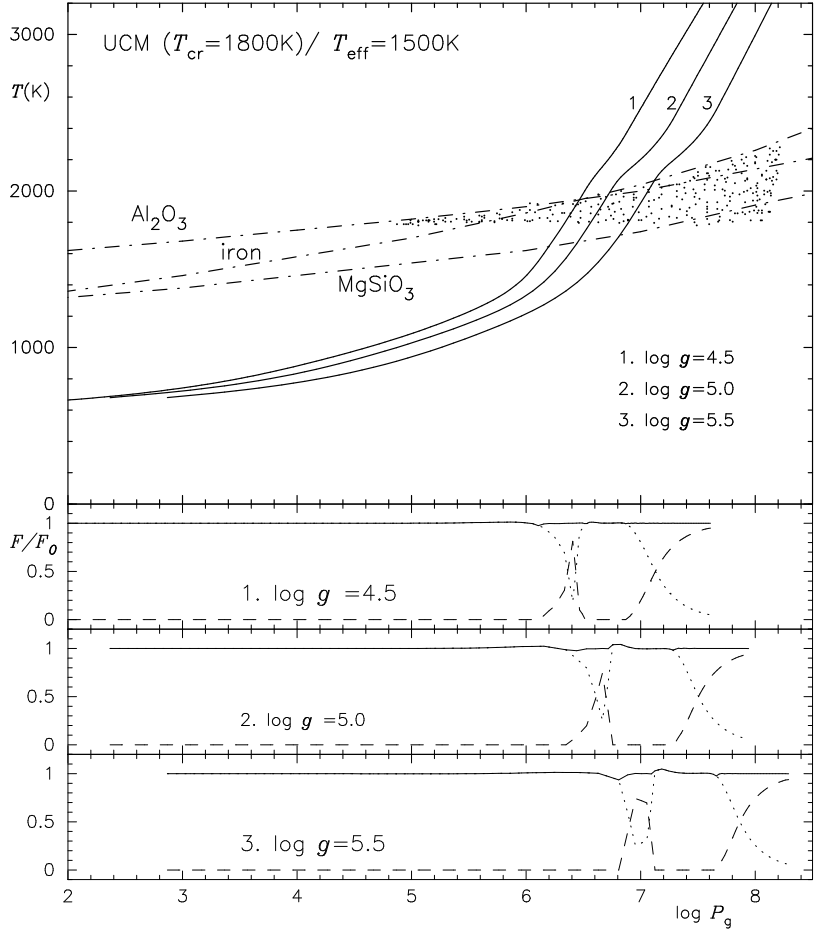}
\caption {
The unified cloudy models  of $T_{\rm eff}$ = 1500K are shown in the 
upper panel for three values of the surface gravities; log $g$ = 4.5, 5.0 
and 5.5 ($v_{\rm micro}=$1 km s$^{-1}$ and the solar metallicity). The 
dot-dashed curves are the dust condensation lines for corundum, iron, and 
enstatite.  The lower three panels show the  radiative, convective, and 
total fluxes normalized by $\sigma T_{\rm eff}^4/\pi$ by the dotted, dashed, 
and solid lines, respectively, for three values of log $g$. 
}
\label{Fig1}
\end{figure}

\begin{figure}
\epsscale{0.9}
\plotone{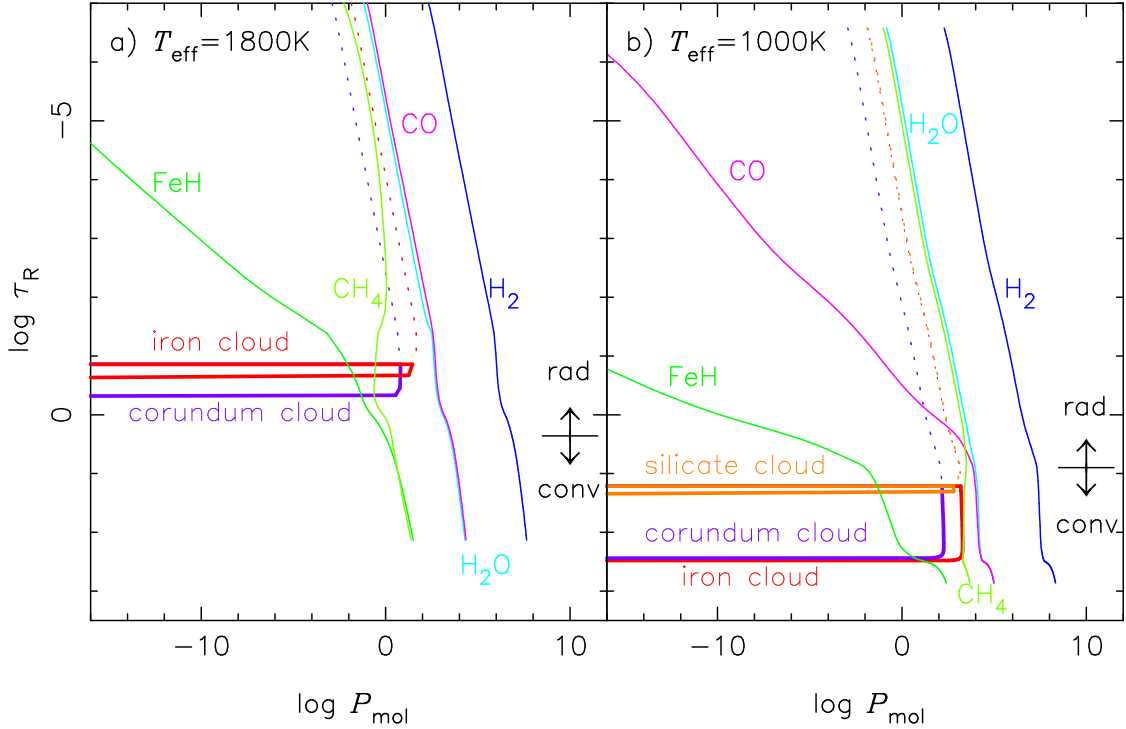}
\caption {
a) Logarithms of the partial pressures (dyn\,cm$^{-2}$) of some molecules 
(abscissa) are plotted against
log\,$\tau_{\rm R}$ (ordinate) in the UCM ($T_{\rm cr} = 1800$\,K) of
$T_{\rm eff} = 1800$\,K and log\,$g$ =5.0. A thin iron cloud and a 
geometrically thicker cloud of corundum are formed in the optically thin 
region. The abundances 
of the dust  grains are shown by the fictitious pressures of  nuclei of the 
refractive elements locked in the dust grains (i.e. Fe and Al for iron and 
corundum, respectively).
The dust abundances under the strict thermodynamical equilibrium are shown 
by the dashed lines. The radiative and convective regimes are indicated.
b) The same for the UCM of $T_{\rm eff} = 1000$\,K and log\,$g$ =5.0.
The iron, corundum, and silicate clouds are formed but in the optically 
thick region.
}
\label{Fig2}
\end{figure}

\begin{figure}
\epsscale{0.65}
\plotone{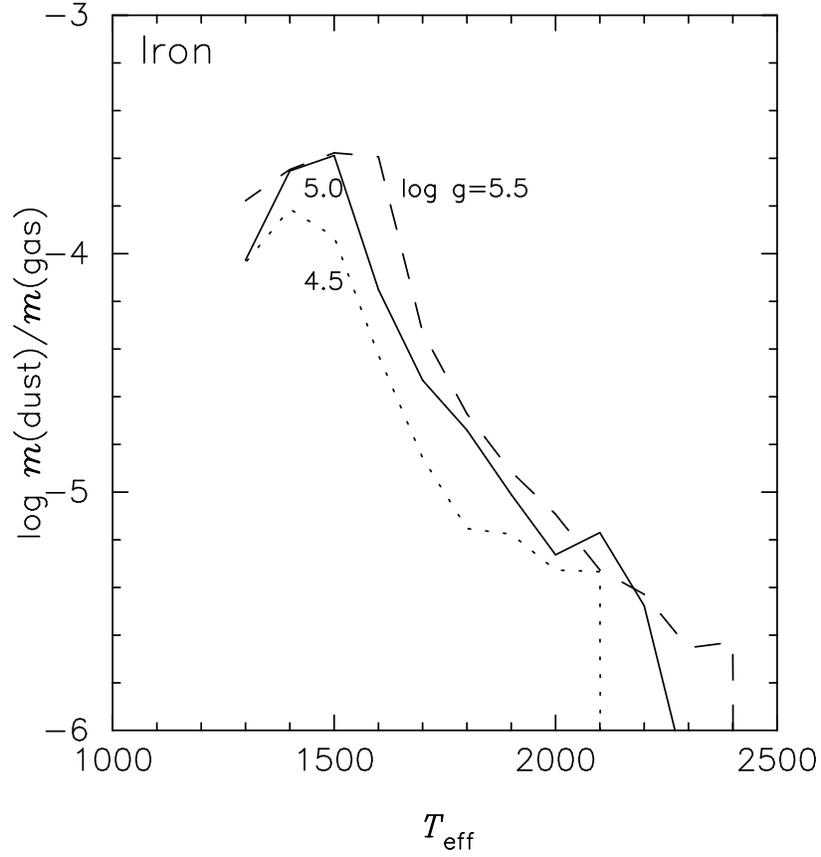}
\caption {
The ratio of the dust mass column density $m$(dust) to the total gas mass 
density $m$(gas)
in the observable photosphere ($\tau_{\rm R} \la 3$ in UCMs with
$T_{\rm cr} = 1800$\,K) is shown in
logarithmic scale against $T_{\rm eff}$ for log\,$g$ = 4.5, 5.0, and 5.5
by dotted, solid, and dashed lines, respectively.
}
\label{Fig3}
\end{figure}

\begin{figure}
\epsscale{0.6}
\plotone{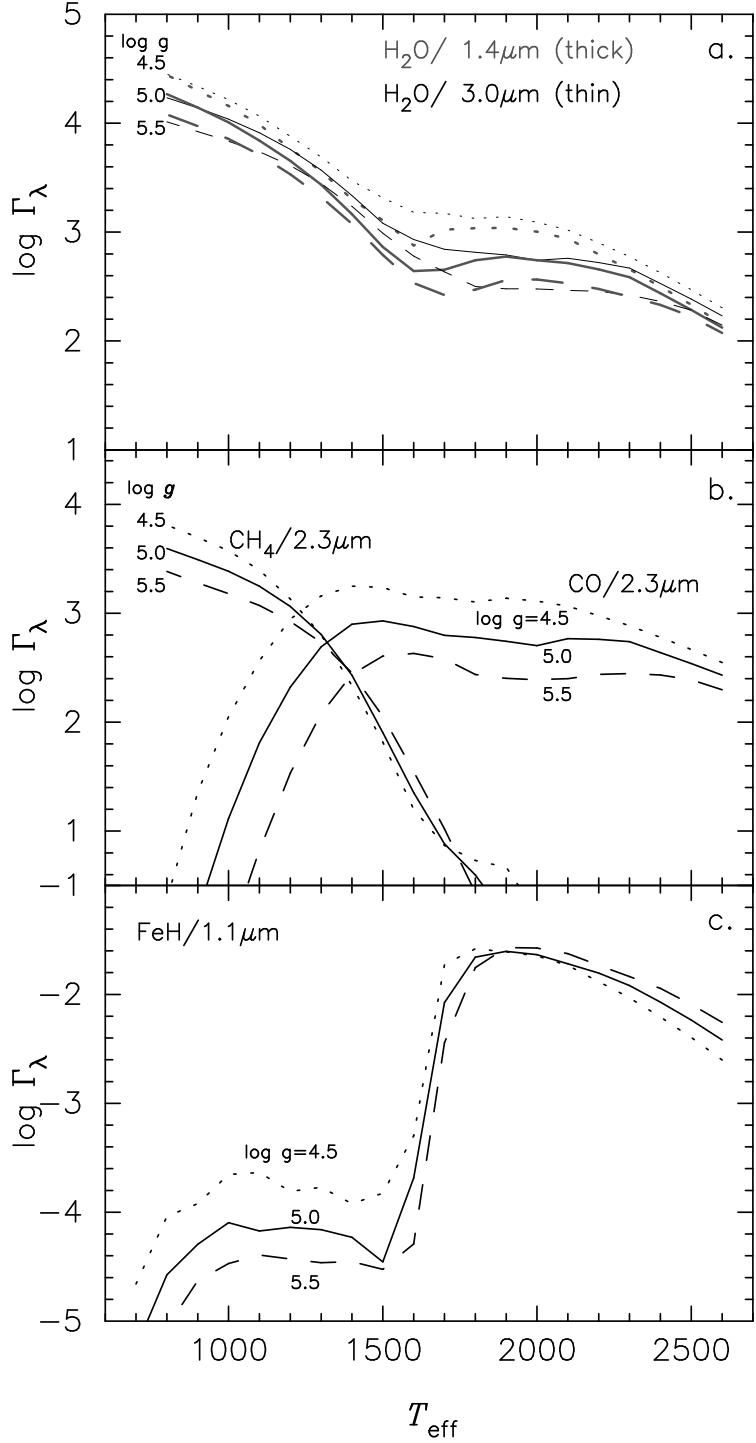}
\caption {
a) The predicted line intensities in logarithmic scale for a line of 
$\chi = 0.0$eV of the H$_{2}$O 1.4 (thick lines) and 3.0\,$\mu$m (thin
lines) bands  plotted against $T_{\rm eff}$  for UCMs ($T_{\rm cr} = 
1800$\,K) with log\,$g$ = 4.5, 5.0, and 5.5 by dotted, solid, and dashed 
lines respectively.
b) The same as for a), but for the CH$_4$ 2.3\,$\mu$m and CO 2.3\,$\mu$m bands.
c) The same as for a), but for the FeH 1.1\,$\mu$m bands.
}
\label{Fig4}
\end{figure}

\begin{figure}
\epsscale{0.65}
\plotone{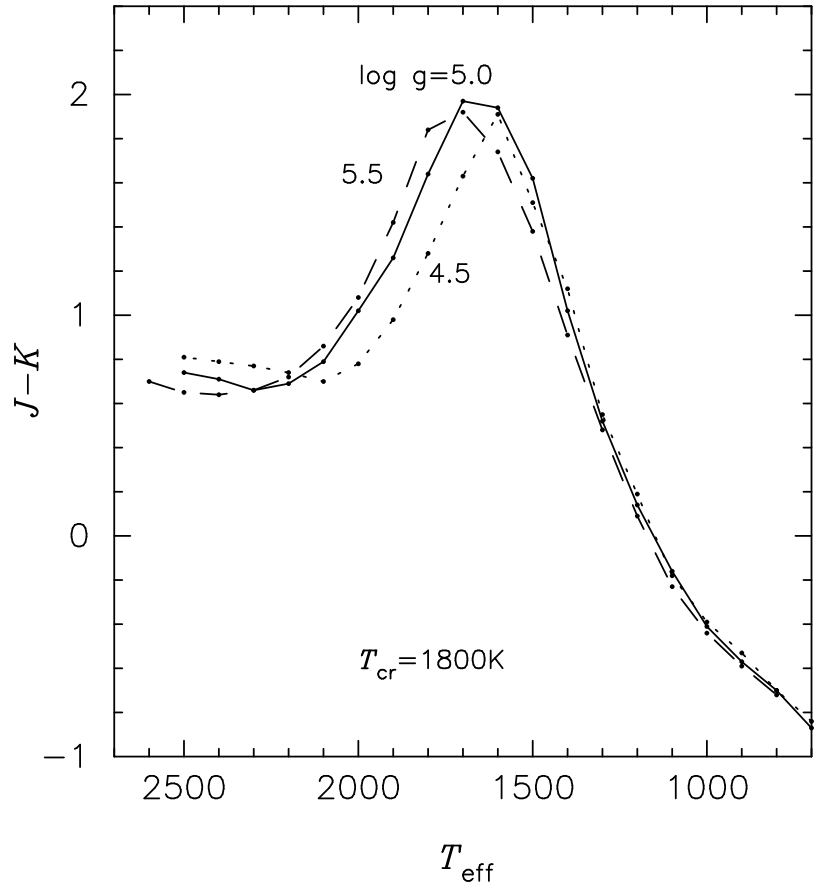}
\caption {
 The predicted  $J-K$ color (MKO system) is plotted against $T_{\rm eff}$  
for UCMs ($T_{\rm cr} = 1800$\,K) with log\,$g$ = 4.5, 5.0, and 5.5
by dotted, solid, and dashed lines, respectively.
}
\label{Fig5}
\end{figure}

\begin{figure}
\epsscale{0.7}
\plotone{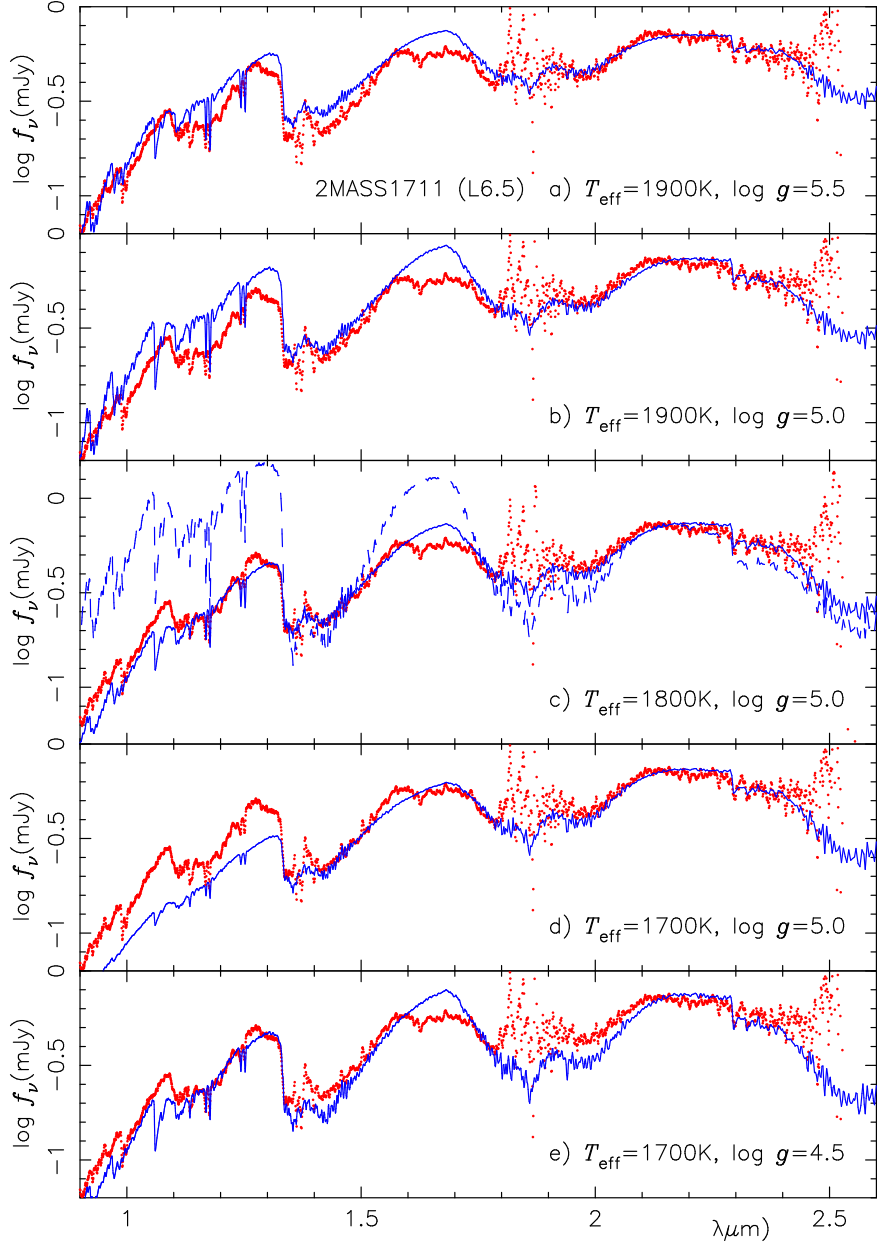}
\caption {
Observed spectrum of the L6.5 dwarf 2MASS\,1711+22 (filled circles) is compared 
with the predicted ones (solid lines) based on the UCMs ($T_{\rm cr} = 
1800$\,K):
a)  $T_{\rm eff} = 1900$\,K and log\,$g$ =5.5.
b)  $T_{\rm eff} = 1900$\,K and log\,$g$ =5.0.
c)  $T_{\rm eff} = 1800$\,K and log\,$g$ =5.0. The dashed line shows
the predicted spectrum based on the model of the same parameters but
all the dust grains are segregated and precipitated below the photosphere
(case C). The difference between the solid and dashed lines indicates the 
effect of the dust clouds.
d)  $T_{\rm eff} = 1700$\,K and log\,$g$ =5.0.
e)  $T_{\rm eff} = 1700$\,K and log\,$g$ =4.5.
}
\label{Fig6}
\end{figure}

\begin{figure}
\epsscale{0.7}
\plotone{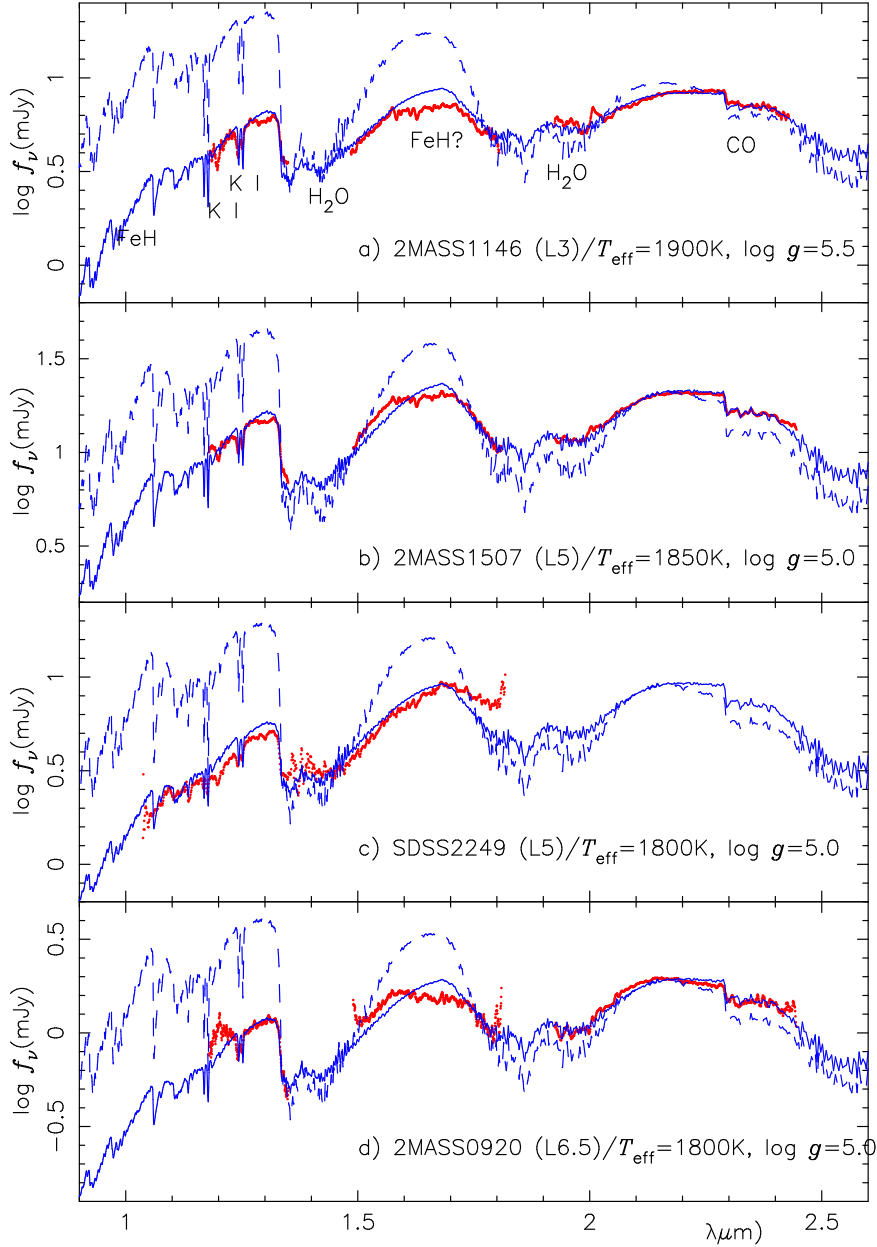}
\caption {
Observed spectra (filled circles) of middle L dwarfs are compared with the
predicted ones (solid lines) based on UCMs ($T_{\rm cr} = 1800$\,K). The
dashed lines have the same meaning as that in Fig.\,6c:
a)  2MASS\,1146+22 (L3) vs. UCM with $T_{\rm eff} = 1900$\,K and log\,$g$ =5.5.
b)  2MASS\,1507-16 (L5) vs. UCM with $T_{\rm eff} = 1850$\,K and log\,$g$ =5.0.
c)  SDSS\,2249+00 (L5)  vs. UCM with $T_{\rm eff} = 1800$\,K and log\,$g$ =5.0.
d)  2MASS\,0920+35 (L6.5) vs. UCM with $T_{\rm eff} = 1800$\,K and log\,$g$ =5.0.
}
\label{Fig7}
\end{figure}

\begin{figure}
\epsscale{0.7}
\plotone{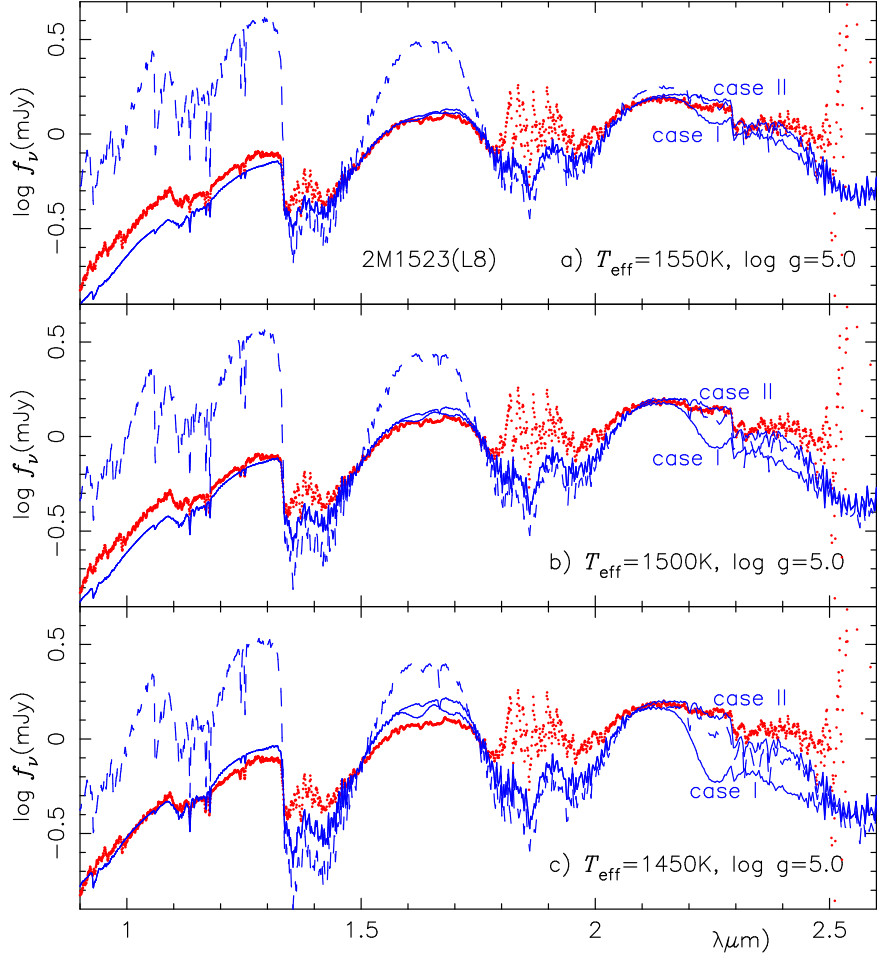}
\caption {
Observed spectrum (filled circles) of the late L dwarf 2MASS\,1523+30 (L8) is 
compared with the predicted ones (solid lines) based on UCMs ($T_{\rm cr} 
= 1800$\,K) of:
a)   $T_{\rm eff} = 1550$\,K and log\,$g$ =5.0,
b)   $T_{\rm eff} = 1500$\,K and log\,$g$ =5.0, and
c)   $T_{\rm eff} = 1450$\,K and log\,$g$ =5.0.
Note that the predicted spectra show bifurcations in the region of 
methane bands near 1.6 and 2.2\,$\mu$m according as the cases I (band model
opacity) or II (linelist) opacities are used for CH$_4$. The dashed lines
have the same meaning as that in Fig.\,6c and only the results based on the 
case II opacity are shown.
}
\label{Fig8}
\end{figure}

\begin{figure}
\epsscale{0.7}
\plotone{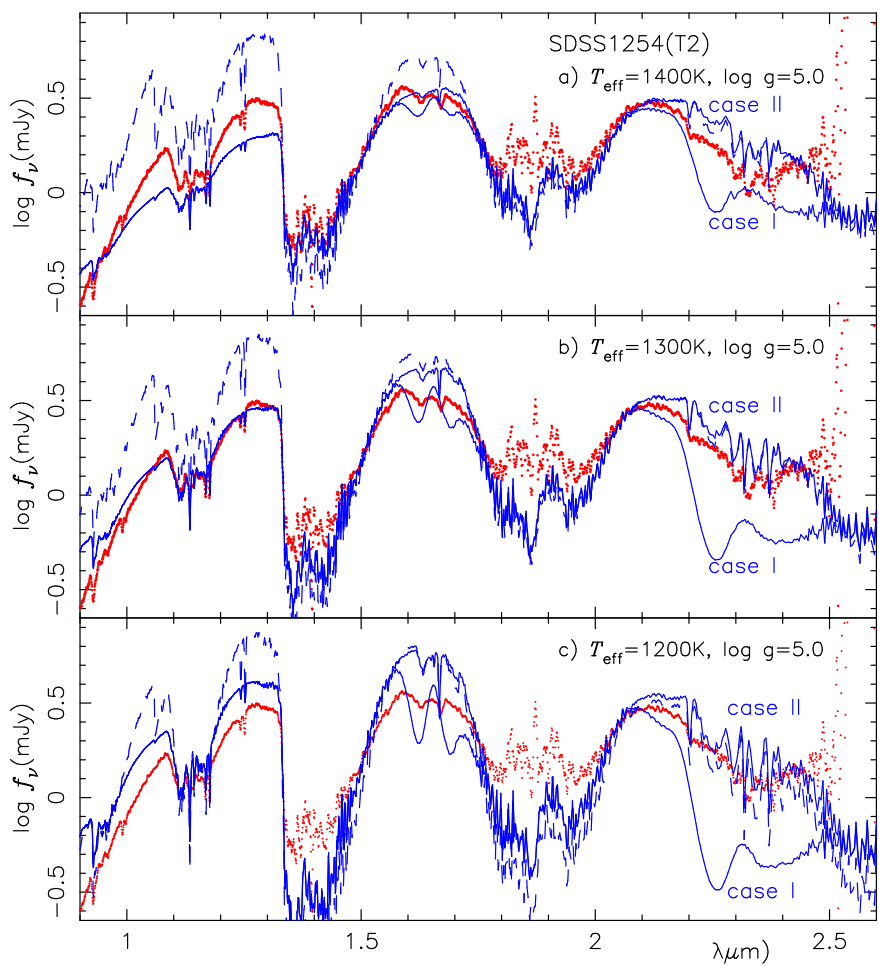}
\caption {
Observed spectrum (filled circles) of the early T dwarf SDSS\,1254-01 (T2) is 
compared with the predicted ones (solid lines) based on UCMs ($T_{\rm cr} 
= 1800$\,K) of:
a)   $T_{\rm eff} = 1400$\,K and log\,$g$ =5.0,
b)   $T_{\rm eff} = 1300$\,K and log\,$g$ =5.0, and
c)   $T_{\rm eff} = 1200$\,K and log\,$g$ =5.0.
See the legend of Fig.8 as for dashed lines and for bifurcations of the 
solid lines. 
}
\label{Fig9}
\end{figure}

\begin{figure}
\epsscale{0.7}
\plotone{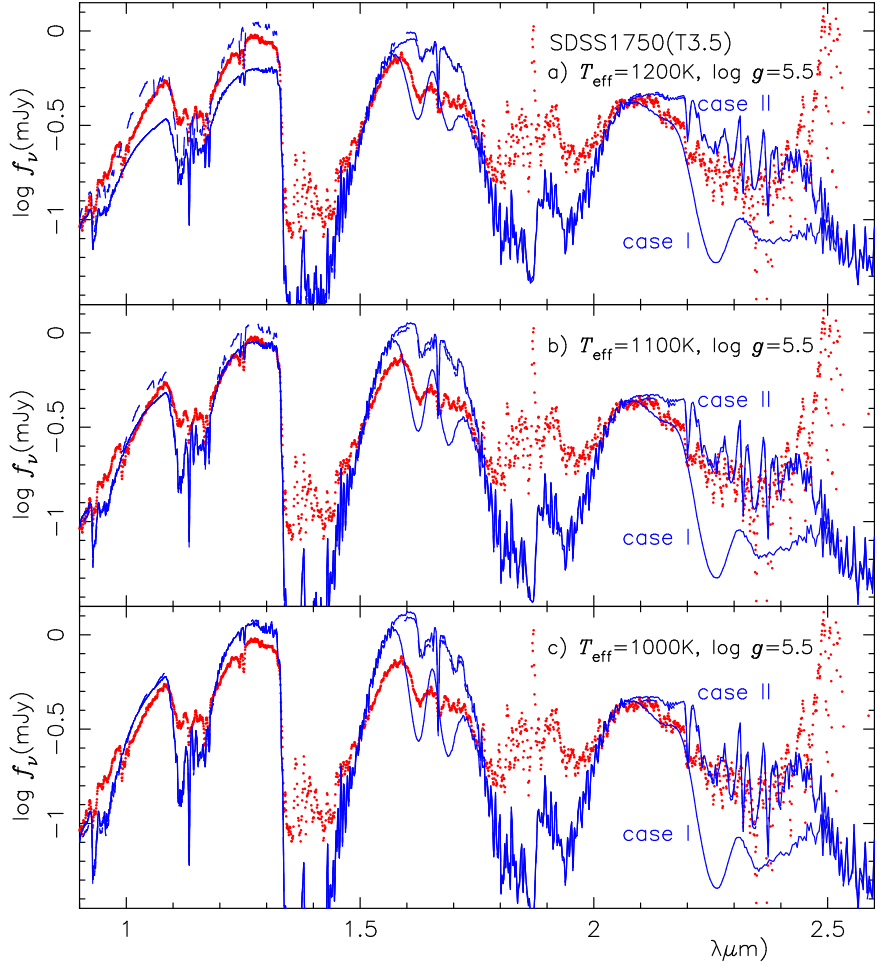}
\caption {
Observed spectrum (filled circles) of the middle T dwarf SDSS\,1750+17 (T3.5) 
is compared with the
predicted ones (solid lines) based on UCMs ($T_{\rm cr} = 1800$\,K) of:
a)   $T_{\rm eff} = 1200$\,K and log\,$g$ =5.5,
b)   $T_{\rm eff} = 1100$\,K and log\,$g$ =5.5, and 
c)   $T_{\rm eff} = 1000$\,K and log\,$g$ =5.5.
The solid lines are now closer to the dashed lines showing the predicted 
spectra from the cloud cleared models, and this fact implies that the
effect of the dust clouds is diminishing according as the clouds
are immersing deeper in the photospheres.
}
\label{Fig10}
\end{figure}

\begin{figure}
\epsscale{0.7}
\plotone{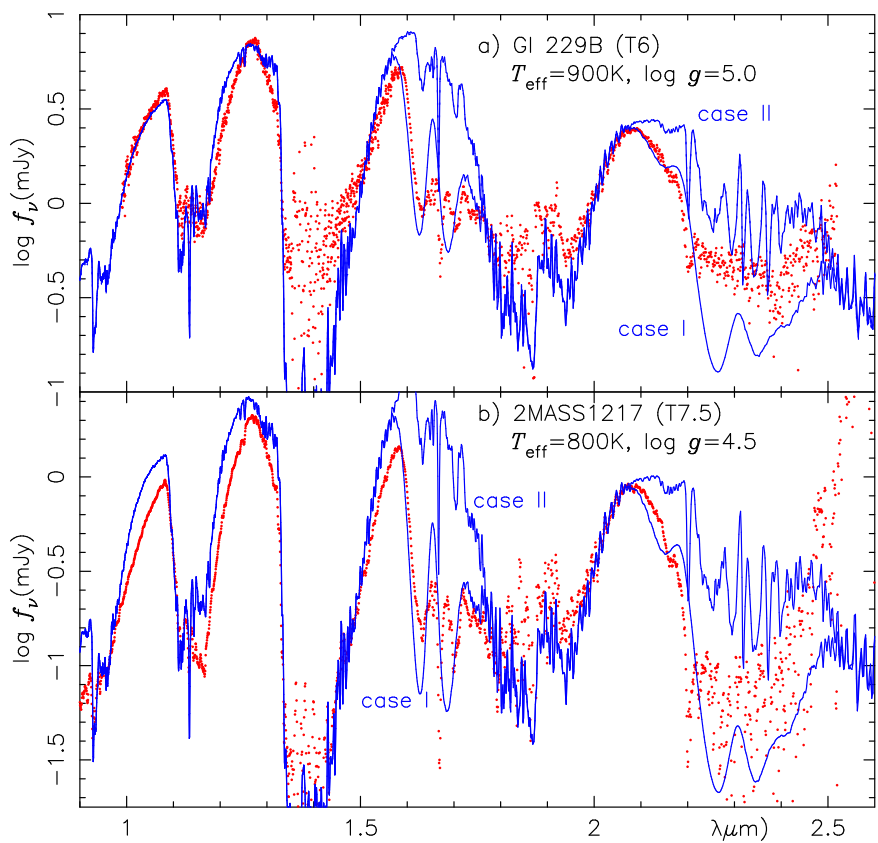}
\caption {
Observed spectra (filled circles) of late T dwarfs are  compared with the
predicted ones (solid lines) based on UCMs ($T_{\rm cr} = 1800$\,K):
a) Gl229B (T6) vs. predicted spectrum by the model of  $T_{\rm eff} = 900$\,K 
and log\,$g$ =5.0.
b) 2MASS\,1217-03 (T7.5) vs. predicted spectrum by the  model of $T_{\rm eff} 
= 800$\,K 
and log\,$g$ = 4.5.
Note that the dashed and solid lines are almost overlapping and this means
that there is almost no effect of the dust clouds on the emergent spectra 
predicted from the UCMs. 
}
\label{Fig11}
\end{figure}

\begin{figure}
\epsscale{0.7}
\plotone{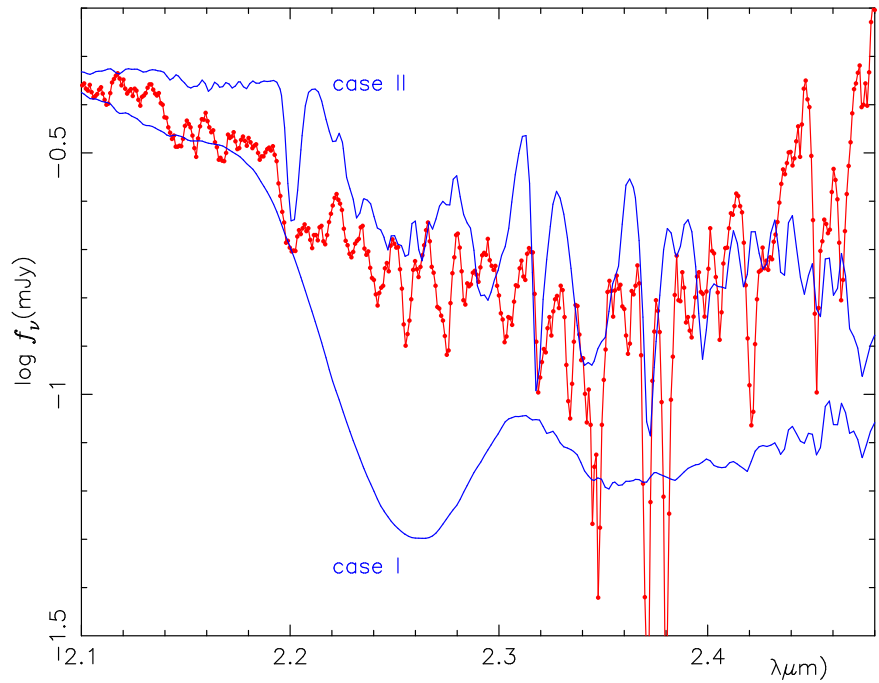}
\caption {
Some details of the
observed methane spectrum of the T3.5 dwarf SDSS\,1750+17 (filled circles 
connected by the solid line) compared with the
predicted ones (solid lines) based on UCMs ($T_{\rm cr} = 1800$\,K) of
   $T_{\rm eff} = 1100$\,K and log\,$g$ =5.5 by the use of the cases I
(band model)  and II (linelist) methane opacities. 
}
\label{Fig12}
\end{figure}

\begin{figure}
\epsscale{0.7}
\plotone{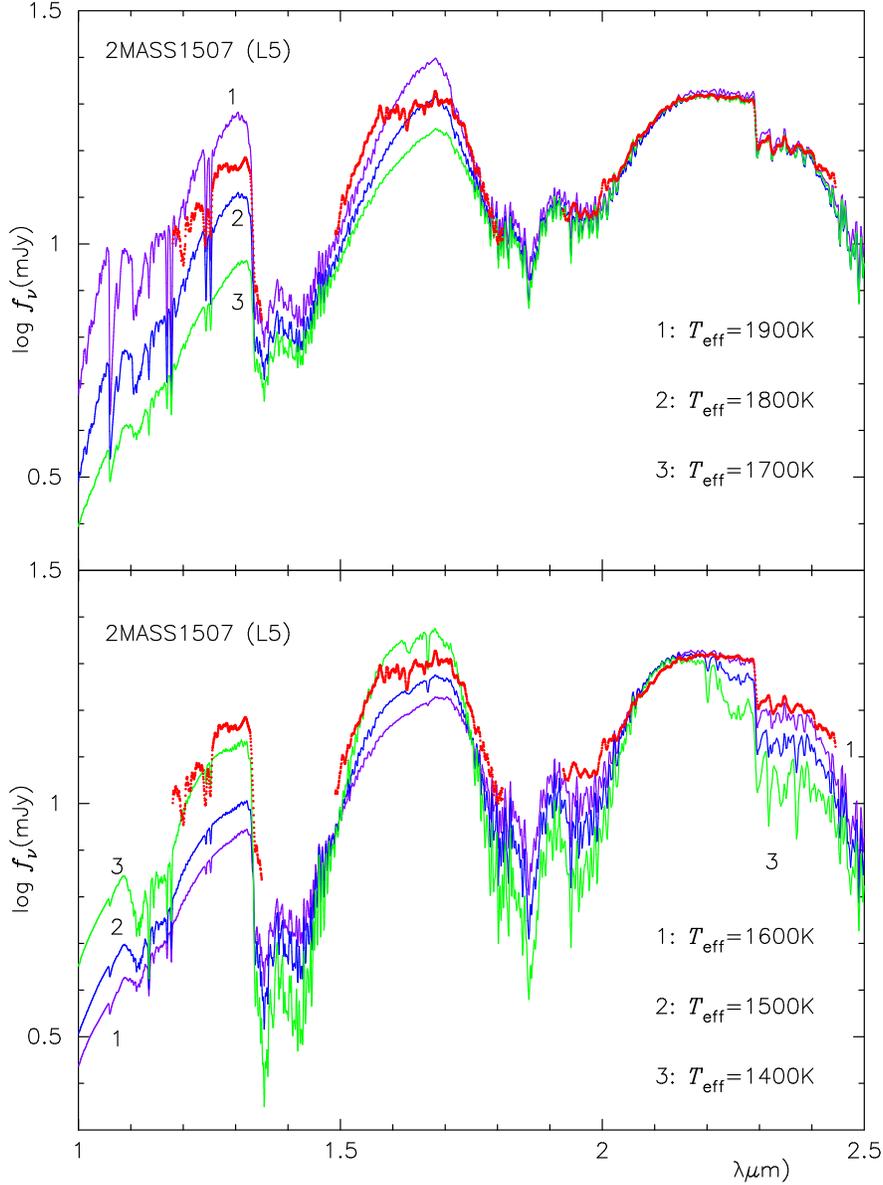}
\caption {
Possible two solutions for the spectrum of the L5 dwarf 2MASS\,1507-16 (filled
circles) are examined by the comparisons with the predicted spectra (solid
lines) of: a) Some  relatively warm models with $T_{\rm eff} = 1700 -
1900$\,K (log\,$g$ = 5.0). The warmer models show  less depression of the 
$J$ flux by the dust extinction because the dust column density is still 
not so large and  we suggested $T_{\rm eff} \approx 1850$\,K (Fig.7b).
b) Some relatively cool models with $T_{\rm eff} = 1400 - 1600$\,K (log\,$g$
= 5.0). The cooler models show  less depression of the $J$ flux by the
dust extinction because of the increased molecular gas above the clouds
which are gradually immersing to the invisible region in these $T_{\rm eff}$
range. Although the overall SED can be fitted with the model of
$T_{\rm eff} \approx 1400$\,K, this models shows too strong methane bands
to be matched with observation, and  this solution cannot be accepted.
}
\label{Fig13}
\end{figure}

\end{document}